\documentclass{aa}  
\usepackage{amsmath,amsfonts,amssymb}
\usepackage{xcolor}
\usepackage{url}
\usepackage[varg]{txfonts}
\usepackage{graphicx}
\usepackage{txfonts}
\usepackage{stfloats}
\usepackage{float}
\usepackage{multirow}
\usepackage{ulem}

\usepackage{natbib,twoopt}
\usepackage[breaklinks=true]{hyperref} %% to avoid \citeads line fills
\bibpunct{(}{)}{;}{a}{}{,} %% natbib format for A&A and ApJ
\makeatletter
\newcommandtwoopt{\citeads}[3][][]{\href{http://adsabs.harvard.edu/abs/#3}%
{\def\hyper@linkstart##1##2{}%
\let\hyper@linkend\@empty\citealp[#1][#2]{#3}}}
\newcommandtwoopt{\citepads}[3][][]{\href{http://adsabs.harvard.edu/abs/#3}%
{\def\hyper@linkstart##1##2{}%
\let\hyper@linkend\@empty\citep[#1][#2]{#3}}}
\newcommandtwoopt{\citetads}[3][][]{\href{http://adsabs.harvard.edu/abs/#3}%
{\def\hyper@linkstart##1##2{}%
\let\hyper@linkend\@empty\citet[#1][#2]{#3}}}
\newcommandtwoopt{\citeyearads}[3][][]%
{\href{http://adsabs.harvard.edu/abs/#3}
{\def\hyper@linkstart##1##2{}%
\let\hyper@linkend\@empty\citeyear[#1][#2]{#3}}}
\makeatother

\usepackage{color}
\hypersetup{colorlinks=true,linkcolor=blue,citecolor=blue,urlcolor=blue}

\usepackage[breaklinks=true]{hyperref} 

\begin{document} 

   \title{Mitigating flicker noise in high-precision photometry}
   \subtitle{I - Characterization of the noise structure, impact on the inferred transit parameters, and predictions for CHEOPS observations}

   \author{S. Sulis \inst{1,2}
          \and M. Lendl \inst{1,3}
          \and S. Hofmeister \inst{4}
          \and A. Veronig \inst{4,5}
          \and L. Fossati \inst{1}
          \and P. Cubillos \inst{1}
          \and V. Van Grootel \inst{6}
          }
          
   \institute{
   Space Research Institute, Austrian Academy of Sciences, 
Schmiedlstr. 6, 8042 Graz, Austria\label{inst2}\label{inst1}
                        \\
                        \email{sophia.sulis@lam.fr}  
                        \and   
Aix Marseille Univ, CNRS, CNES, LAM, Marseille, France
             \and
            Observatoire de l'Université de Genève, 51 chemin des Maillettes, 1290 Sauverny, Switzerland \label{inst3} 
             \and
             Institute of Physics, University of Graz, Universit\"atsplatz 5, 8010 Graz, Austria \label{inst4}
                        \and
             Kanzelh\"ohe Observatory for Solar and Environmental Research, University of Graz, Kanzelh\"ohe 19, 9521 Treffen, Austria \label{inst5}
            \and
             Space sciences, Technologies and Astrophysics Research (STAR) Institute, Université de Liège, 19C Allée du
             six-ao{\^u}t, B-4000 Liège, Belgium \label{inst6}
             }

   \date{XXX}
 
%%%%%%%%%%%%%%%%%%%%%%%%%%%%%%%%%%%%%%%%
% ABSTRACT AND KEYWORDS
%%%%%%%%%%%%%%%%%%%%%%%%%%%%%%%%%%%%%%%%
\abstract
%% context
{In photometry, the short-timescale stellar variability (``flicker''), such as that caused by granulation and solar-like oscillations, can reach amplitudes comparable to the transit depth of Earth-sized planets and is correlated over the typical transit timescales.
It can introduce systematic errors on the inferred planetary parameters when a small number of transits are observed.}
%% aims
{The objective of this paper is to characterize the statistical properties of the flicker noise and quantify its impact on the inferred transit parameters.}
%% methods
{We used the extensive solar observations obtained with SoHO/VIRGO to characterize flicker noise. 
We simulated realistic transits across the solar disk using SDO/HMI data and used these to obtain transit light curves, which we used to estimate the errors made on the transit parameters due to the presence of real solar noise. We make these light curves publicly available. 
To extend the study to a wider parameter range, we derived the properties of flicker noise using Kepler observations and studied their dependence on stellar parameters. 
Finally, we predicted the limiting stellar apparent magnitude for which the properties of the flicker noise can be extracted using high-precision CHEOPS and PLATO observations.}
%% results
{Stellar granulation is a stochastic colored noise, and is stationary with respect to the stellar magnetic cycle. 
Both the flicker correlation timescales and amplitudes increase with the stellar mass and radius.
If these correlations are not taken into account when fitting for the parameters of transiting exoplanets, this can bias the inferred parameters. In particular, we find errors of up to $10\%$ on the ratio between the planetary and stellar radius ($R_p/R_s$) for an Earth-sized planet orbiting a Sun-like star. }
%% conclusions 
{Flicker will significantly affect the inferred parameters of 
transits observed at high precision with CHEOPS and PLATO for F and G stars. Dedicated modeling strategies need to be developed to accurately characterize both the star and the transiting exoplanets.}

% Put max 6 keywords
   \keywords{ 
   <  Planetary systems - Techniques: photometric  - Stars: activity - Sun: 
granulation - Methods: statistical >
   }
   \maketitle
   
%%%%%%%%%%%%%%%%%%%%%%%%%%%%%%%%%%%%%%%%
% INTRODUCTION
%%%%%%%%%%%%%%%%%%%%%%%%%%%%%%%%%%%%%%%%

\section{Introduction}

The following decade will see the outcome of several missions in the field of extrasolar planets. With the new space missions like the \textit{Transiting  Exoplanet  Survey  Satellite} (TESS; \citeads{2015JATIS...1a4003R}), the \textit{Characterizing Exoplanet Satellite} (CHEOPS; \citeads{2014SPIE.9143E..2JF}), and the \textit{Planetary Transits and Oscillations of stars} mission (PLATO; \citeads{2014ExA....38..249R}), we expect to be able to detect and precisely characterize several thousands of new transiting Neptune- to Earth-like planets. 
However, as it was already the case with bright stars observed with Kepler \citepads{2011ApJS..197....6G,2015AJ....150..133G}, the high-precision photometry of these instruments will not be limited by photon noise but by stellar variability. Indeed, at the stellar surfaces, several phenomena (e.g., spots, plages, flares, convection, oscillations) evolve on different timescales and generate variability that degrades the detection and the shape determination of planetary transits (see e.g., \citeads{2018ASSP...49..239O} for a recent review). 

In this paper, we focus on the stellar variability taking place on timescales similar to the duration of a single planetary transit (i.e., $<1$ day). On these timescales, the dominant stellar activity contribution of quiet stars of around solar mass comes from the surface convective motions and the pressure-mode oscillations. 
In the case of exoplanet transits, star spot crossings can also punctually contribute to the short-timescale noise. In this study however, we disregard these punctual noise sources as they are not regularly present in the observations, and their contribution can be averaged out by analyzing hundreds of 
short individual time series.

Oscillations are studied extensively through asteroseismology. Pressure-mode oscillations are generated in the stellar convective envelope of stars of solar mass and allow us to probe the stellar interiors. Their characteristics (amplitudes and periods) directly inform on the evolutionary stage of the star. Solar-type stars oscillate with a period of several minutes and generate a photometric background signal with an amplitude of several tens of parts per million (ppm; \citeads{1988IAUS..123..497H}). The amplitudes and frequencies of these p-modes are known to change with the stellar magnetic cycle \citepads{2011ApJ...732L...5C,Salabert_2011, 2013JPhCS.440a2020G}.

Convection results from turbulent plasma motion at the stellar surface. When resolved, ascending hot plasma surfacing at the granules appears bright, and cool plasma descending in the intergranular lanes appears darker, leading to variable contrasts in brightness over time. 
The individual granulation cells can only be resolved for the solar surface, where the individual cells have average sizes of $1000$ km \citepads{1989ApJ...336..475T} and a median ``turnover'' timescale between $7$ and $10$ minutes \citepads{Nesis2002}. 
However, granulation is an evolving process: the size of the granules grows and shrinks with time, the cells merge and split with the surrounding granules, and the resulting photometric variability appears correlated over timescales larger than the turnover period \citepads{2011A&A...532A.108S}.
At large scales ($\sim3\times10^7$ m), we can also observe a conglomerate of convection cells when mapping the  magnetic flow on the solar surface. 

Thanks to the \textit{Solar and Heliospheric Observatory} (SoHO) measurements, the photometric signature of solar granulation is known to be approximately $100$ ppm (\citeads{1988IAUS..132..239D}, \citeads{1997SoPh..170....1F}; 
\citeads{2004A&A...414.1139A}).
The typical amplitudes and turnover timescales of surface granulation depend on the stellar parameters. Amplitudes increase with effective temperature (and possibly the stellar metallicity; see \citeads{2017A&A...605A...3C}) and decrease with decreasing stellar mass and/or increasing surface gravity \citepads{2005ESASP.560..979S}. Both decrease with increasing mean oscillation frequency $\nu_{max}$ \citepads{1988IAUS..132..239D,2014A&A...570A..41K}. Recently, high-precision Kepler measurements have made it possible to measure the photometric amplitude of the granulation variability in increasing detail, which has given rise to a new technique for deriving the stellar surface gravity \citepads{2011ApJ...741..119M, 2016ApJ...818...43B,2014ApJ...781..124C,2018MNRAS.480..467P} and density \citepads{2041-8205-785-2-L32}.

While granulation contains extensive information for stellar physics, because of its stochastic nature it is typically considered as noise from the point of view of detecting and/or characterizing planets and their atmospheres. 
In an exoplanetary context, granulation is often referred to \textit{flicker} noise as its power spectrum follows the form of a power-law function in a specific frequency range related to its turnover timescales. 
In practice, because the granulation turnover timescale is  small ($<<1$ day) for solar-type stars, this stellar variability can be considered uncorrelated from one planetary transit to another. Therefore, the influence of this noise can be reduced when several transit events are observed and the overall signal-to-noise ratio (S/N) can be improved by phase-folding the planetary transits with the planet orbital period.
However, this technique is not efficient when only a small number of transits are observed (e.g., for long period planets) or when the transits show transit timing variations (multiplanetary systems).
Hence, the statistical properties of the flicker noise need to be known in order to design dedicated modeling strategies correctly accounting for (and possibly reducing) its influence when estimating the parameters of transiting exoplanets. 

Granulation noise is often modeled using Harvey law profiles \citepads{1985ESASP.235..199H} with parameters (amplitude and timescales) estimated based on the observed stellar power spectral density (PSD) \citepads{1999ASPC..173..297P,2014A&A...570A..41K,2014ApJ...781..124C}. One success of this technique is the relation between the parameters of the Harvey law profiles and the stellar fundamental parameters, which appear to be correlated \citepads{2014A&A...570A..41K}.   
Recently, \citetads{2019MNRAS.489.5764P} proposed to couple Gaussian Processes (GP) noise modeling \citepads{Rasmussen:2005:GPM:1162254} with PSDs based on Harvey-like profiles and showed that taking into account the flicker correlations within the transit analysis tends to improve the accuracy of the inferred transit parameters. \citetads{2015ApJ...800...46B} came to a similar conclusion using other GP noise models to fit for granulation and derive the parameters of the transiting hot Jupiter Kepler-91b.
A more realistic way to model stellar granulation is to turn to modern three-dimensional radiative hydrodynamical simulations of stellar convection \citepads{2009LRSP....6....2N}. While computationally demanding, these codes allow the generation of realistic photometric time series of flicker noise and could provide valuable diagnostics to extract the flicker properties affecting transits \citepads{2017A&A...597A..94C}. This has been shown by \citetads{2015A&A...576A..13C} with the transit of Venus in 2004.

In this work, we aim to understand the effect of stellar flicker on determining accurate and precise properties of exoplanets through transit photometry and describe how this noise behaves as a function of the stellar parameters. This paper is organized as follows. In Sec.~\ref{Sec2}, we summarize the statistical properties of solar flicker using $21$ years of continuous solar observations provided by SoHO. In Sec.~\ref{Sec3}, we analyze the impact of solar flicker on the determination of the transit parameters. In Sec.~\ref{Sec4}, we 
use short-cadence Kepler observations to extract the flicker properties and show their dependence with the stellar parameters. In Sec.~\ref{Sec5}, we discuss the potential for the photometric characterization of this noise source with the future high-precision observations of CHEOPS and PLATO. We conclude in Sec.~\ref{ccl}.

%%%%%%%%%%%%%%%%%%%%%%%%%%%%%%%%%%%%%%%%
% SEC.II: VIRGO DATA ANALYSIS
%%%%%%%%%%%%%%%%%%%%%%%%%%%%%%%%%%%%%%%%

\section{Statistical characterization of short-timescale solar variability}
\label{Sec2}

This section summarizes the statistical properties of solar granulation in photometric observations. We aim to list the properties that are needed by signal processing routines to analyze the influence of this noise source and derive appropriate noise models.% Several studies have pursued this goal previously, \textcolor{red}{[Add citations]}. 

\subsection{VIRGO observations}
\label{Sec21}

The Sun has been monitored since 1996 by the ESA/NASA SoHO spacecraft.
Onboard, the \textit{Variability of solar IRradiance and Gravity Oscillations} 
(VIRGO) experiment measures the spectral irradiance with a three-channel 
sun photometer (SPM) of $5$ nm bandwidth at wavelengths of $402$ (blue), $500$ 
(green), and $862$ (red) nm. 
Observations at these wavelengths span different heights in the solar photosphere (from $-20$ km compared to the base of the solar photosphere for the green and blue channels to $+10$ km for the red, \citeads{2005ApJ...623.1215J}).
The data are integrated over $60$ seconds and centered around the full minute.
The duty cycle of the $21$ years of VIRGO time series (from April 11, 1996, to 
March 30, 2017) is around $96\%$, making it the best data set 
available today to derive the statistical properties of the solar variability.  
The photon noise in each of the three channels is below $10$ ppm \citepads{2017A&A...608A..87S}.
A full description of the instrument's characteristics and technical 
calibration procedures can be found in \citetads{1995SoPh..162..101F}, 
\citetads{1997SoPh..170....1F}, and \citetads{2002SoPh..209..247J}.

The available data (level 1) have been converted to physical units (W/m$\rm^2$/nm), corrected by 
temperature variations, and calibrated to a constant distance between the spacecraft and the Sun.
We carefully corrected for the instrumental degradation\footnote{Higher 
instrumental ageing affects the blue channel detector.} evolving over time following the procedure below:
\begin{enumerate}
    \item For computational reasons, we split the whole data set into 
subseries, each spanning $365$ days.
    \item We smoothed each subseries using a running average of $3$ days in length 
and localized the $3\sigma$ outliers. These outliers were disregarded in step 3.
    \item We fitted a high-degree polynomial function to the smoothed time 
series and used it to normalize the initial time series (containing the 
outliers).
    \item We removed the $5\sigma$ outliers from each corrected time series, 
    \item We finally compared the resulting three SPM datasets and kept the common data points between these three series. 
\end{enumerate}
 
\begin{figure}[t] \centering
    \resizebox{\hsize}{!}{\includegraphics{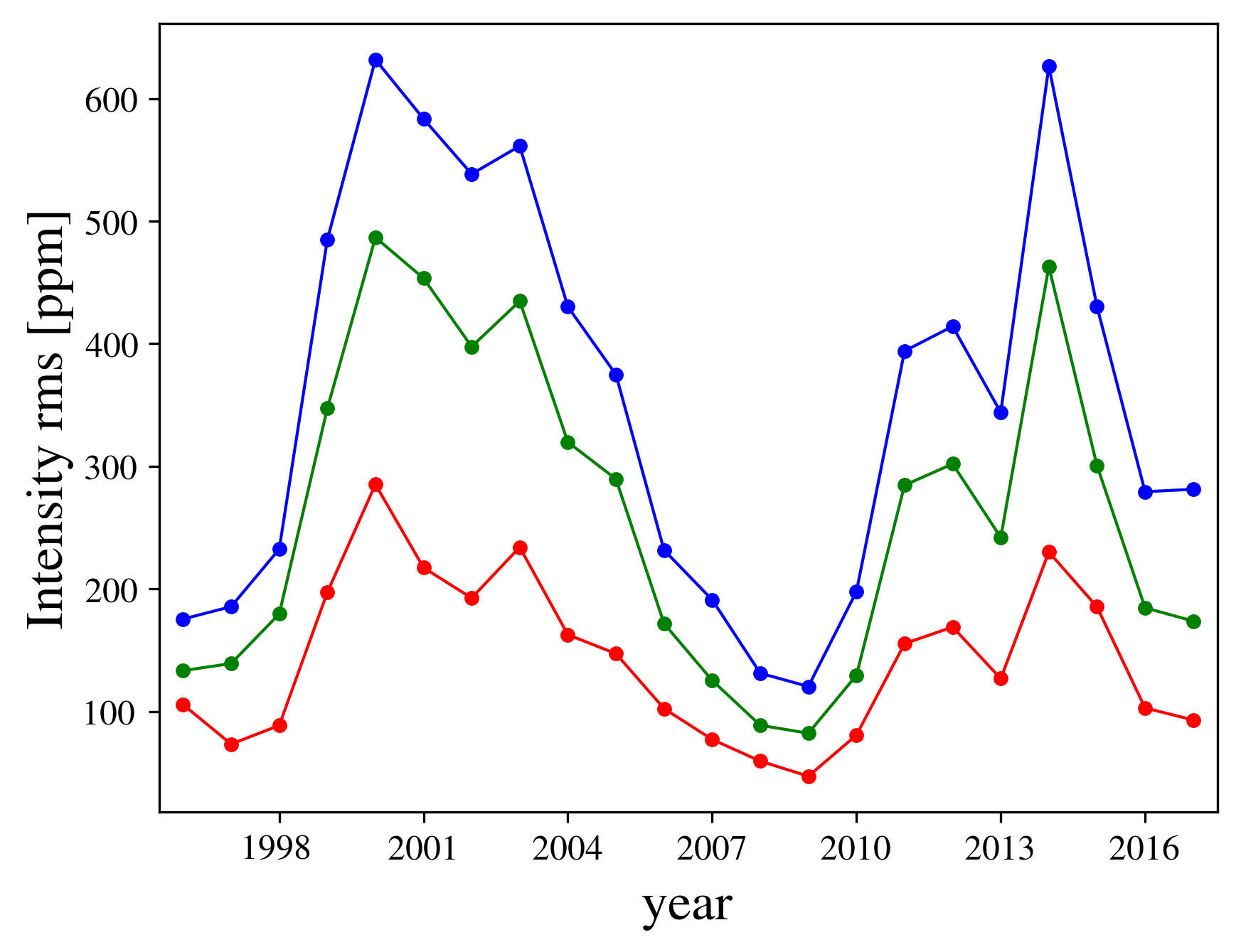}}
    \caption{Yearly RMS of the VIRGO time series for the blue, green, and red SPM channels (see colors). Solar minima at the beginning of cycles 23$^{\rm }$ (from 1996 to 2008) and 24$^{\rm }$ (since 2009)  are clearly visible. }
    \label{Fig_rms}
\end{figure}

As the emergent flux decreases from optical to infrared, we observe a strong color dependence between the datasets, with higher variability at short (blue) wavelengths 
than at long (red) wavelengths (see also \citeads{2003ASPC..294..441A}).
 This is illustrated in Fig.~\ref{Fig_rms} through the time series root-mean-square (RMS) measurement, where the signature of the approximately eleven-year solar cycle is evident. 
To study the short-timescale solar variability, we finally divided the detrended datasets into one-day subseries, and removed those subseries that had missing data and 
strong instrumental features. We obtained a total of $5912$ regularly sampled one-day subseries (i.e., $\approx$ 275 subseries/year). We use this dataset to characterize the properties of the 
granulation activity in the following subsections.

\subsection{A stationary stochastic colored noise}
\label{sec_22}

\begin{figure}[t] \centering
    \resizebox{\hsize}{!}{\includegraphics{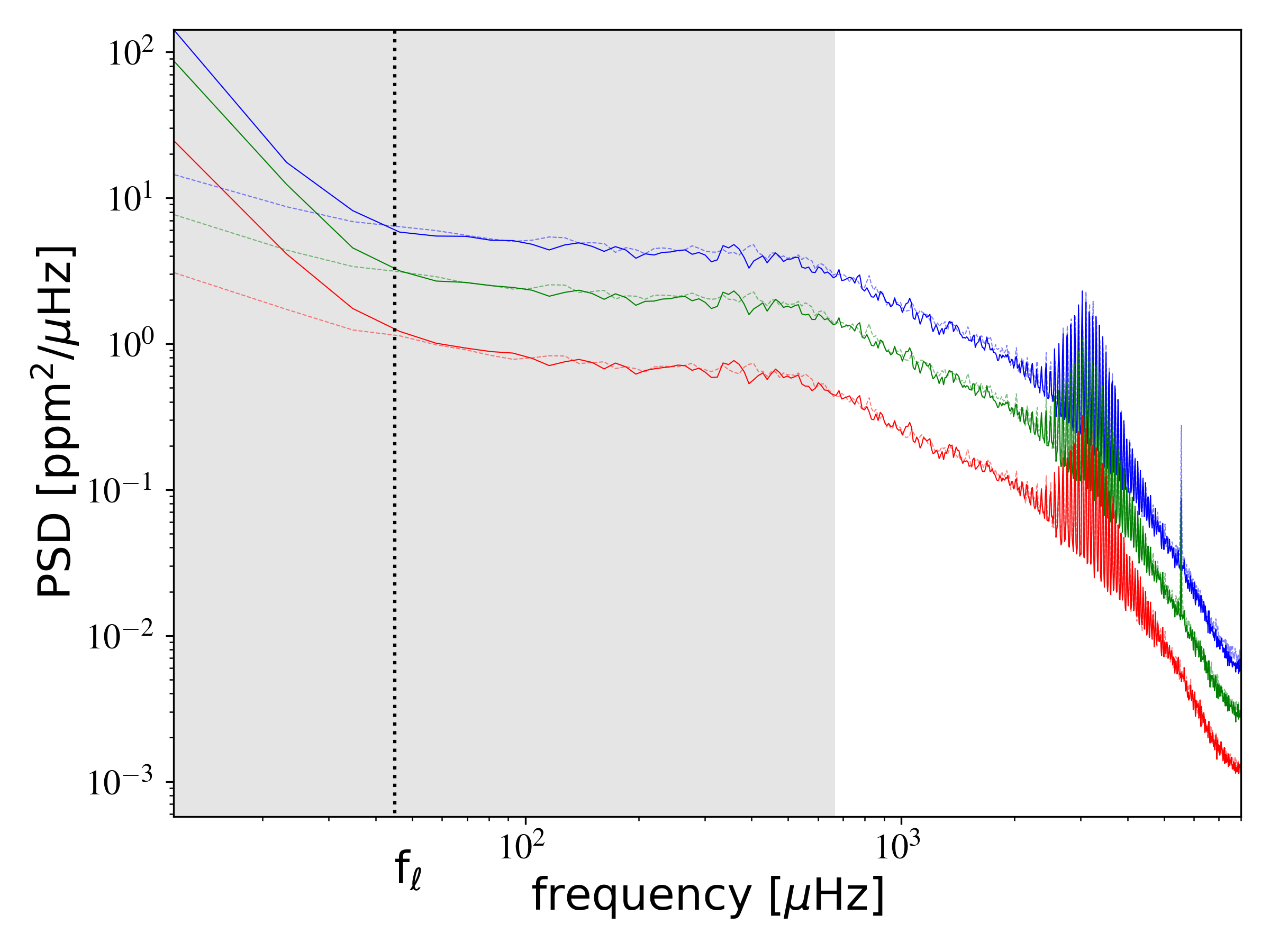}}
    \caption{Averaged periodograms computed with Eq. \eqref{eq_PL} using all one-day subseries available during a minimum (2008, solid dashed lines) and a maximum (2003, solid solid lines) of the solar cycle. Each color indicates the observation of 
one SPM channel. Differences with the solar cycle are observed at frequencies  $\nu < f_{l}$ (i.e., at periods $>6.2$ hr). The peak at $\nu=5570~\mu$Hz affecting the PSD of the blue and green SPM data is an electronic artifact related to the calibration period used by the acquisition system of VIRGO. 
The gray shaded area indicates the frequency range corresponding to the typical duration of planetary transit (from $\sim 25$ min to several hours).  }
    \label{Fig_periodo}
\end{figure}

The solar short-timescale variability (periods $< 1$ day) is dominated by 
instrumental noise, oscillation modes, and convection. 
The distinct signatures of these three noise sources can be well discerned in the PSD of the VIRGO observations. Figure~\ref{Fig_periodo} shows an estimate of this PSD through the averaged periodogram defined as \citepads{10.2307/2332141}\footnote{The definition based on Eq. \eqref{eq_PL} differs slightly from \citetads{10.2307/2332141}, who proposed to split a given time series $X(t)$ into $L$ subseries $X_\ell(t)$. }:
\begin{equation} 
        P_L(\nu_k): =  \frac{1}{L} \sum_{\ell=1}^{L}  \frac{1}{N} \Big| 
\sum_{j=1}^{N}  ~X_\ell(t_j)~\mathrm{e}^{-{{\rm i}}2\pi\nu_k t_j} \Big| ^2,
    \label{eq_PL}
\end{equation}
with $L$ being the number of available one-day subseries, 
$\{X_\ell\}_{\ell=0,\hdots,L}$ , per year, $N$ the number of data points of these subseries, and $t_j = j\times dt$ the time of the measurements with $j=0,\hdots,N.$ Also,  $dt$ is the temporal sampling and $\nu_k:=\frac{k}{N}$ are the Fourier frequencies\footnote{In the 
following, we use the Greek letter ``$\nu$'' for the Fourier frequencies in 
general and the Latin letter ``$f$'' for pointing to a characteristic frequency 
of $\nu_k$ (except for the p-modes frequency, $\nu_{max}$).} computed for $k=0,\hdots,N-1$.
We observe a strong frequency dependence of the PSD due to correlations over several timescales.  This frequency dependence is characteristic of a noise that is defined as a \textit{colored} noise, in opposition to a white Gaussian noise (WGN) that shows a constant PSD over the whole frequency range\footnote{Useful notes: ``Gaussian'' and ``colored'' are unrelated properties. A random process is Gaussian if it is normally distributed (i.e., Gaussian probability distribution function). If the process is white, the covariance matrix is the Identity matrix (and the PSD is constant); if it is colored, the covariance matrix is not Identity (and the PSD is not constant in frequency). (Second-order) Stationarity means that second-order statistics (i.e., the correlation structure) do not change in time.}. 
The periodic signatures of solar oscillations are clearly visible around $3000$ $\mu$Hz. Granulation dominates the frequency range corresponding to a typical transit duration (gray area): between activity and oscillations. As long as the planet transit does not cross a spot or plage, this may be the dominant stellar noise source in high-precision photometric observations.
By comparing the averaged periodograms computed using subseries taken during a minimum and a maximum of the solar cycle (see dashed and solid lines, respectively), we only note significant differences at low frequencies ($\nu < f_{\ell}=44.9 ~\mu$Hz), in agreement with \citetads{2011A&A...532A.108S}. Although not visible in this comparison, we also note the presence of small changes of the p-modes signatures (both in amplitude and frequency) over the magnetic cycle, in line with findings by \citetads{2011ApJ...732L...5C}, \citetads{Salabert_2011}, and \citetads{2013JPhCS.440a2020G}. 
Interestingly, the frequency region that is dominated by the granulation variability,  $\nu \in [f_{\ell},~2300]$ $\mu$Hz, is not significantly affected by the solar magnetic cycle (see \citeads{2011A&A...532A.108S}; \citeads{2018A&A...616A..87M}). Therefore, we can state that granulation variability is stationary with respect to the solar cycle (i.e., its statistical 
properties remain constant over the solar cycle).

\subsection{Time domain analysis: amplitude distributions}

\begin{figure}[t] \centering
    \resizebox{\hsize}{!}{\includegraphics{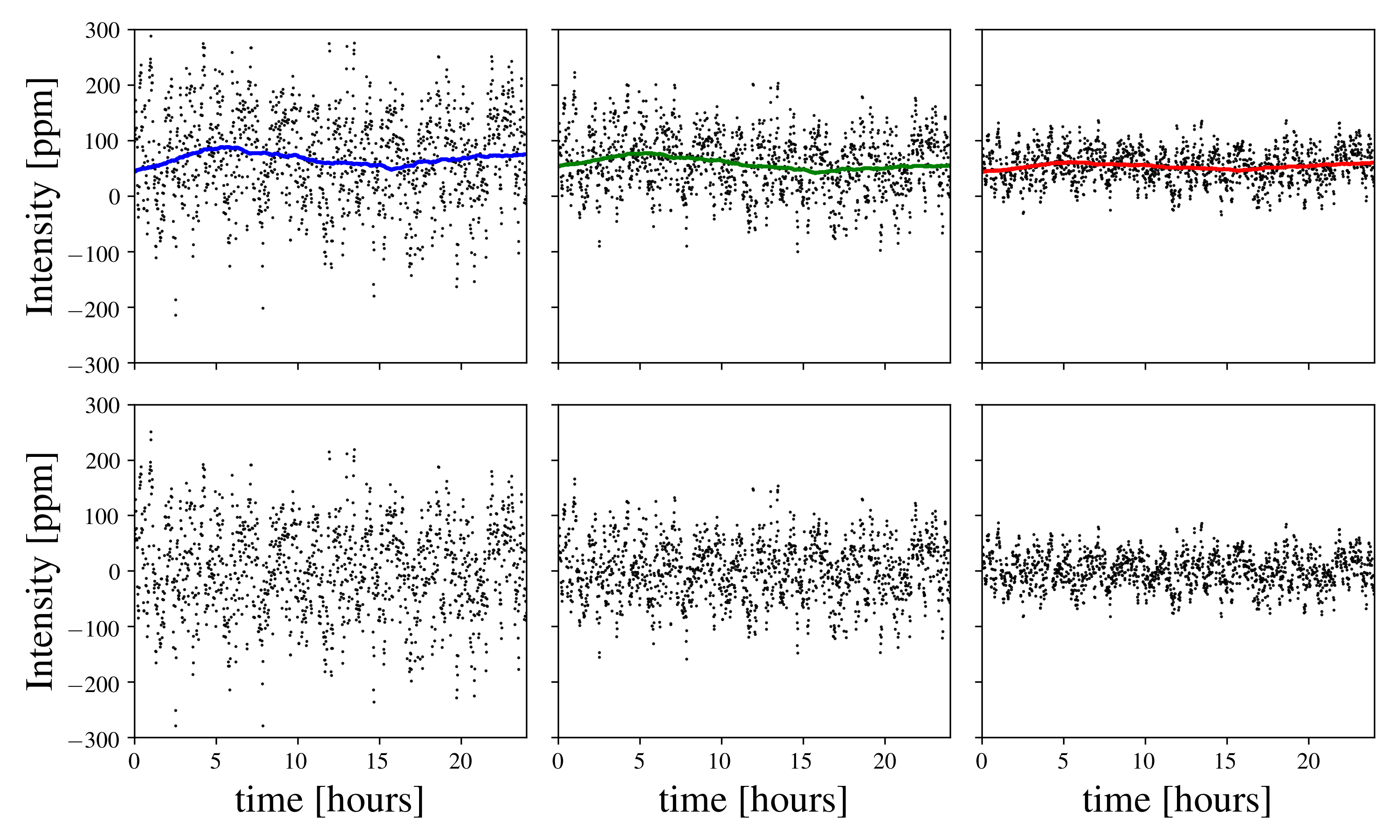}}
    \caption{Example of a one-day VIRGO time series (top) detrended (bottom) using a SG filter with a window size of $15$ hours (colored solid lines). 
From left to right: Dataset from the blue, green, and red SPM channels.}
    \label{Fig_ts}
\end{figure}

\begin{figure*}[b]
    \resizebox{\hsize}{!}{\includegraphics{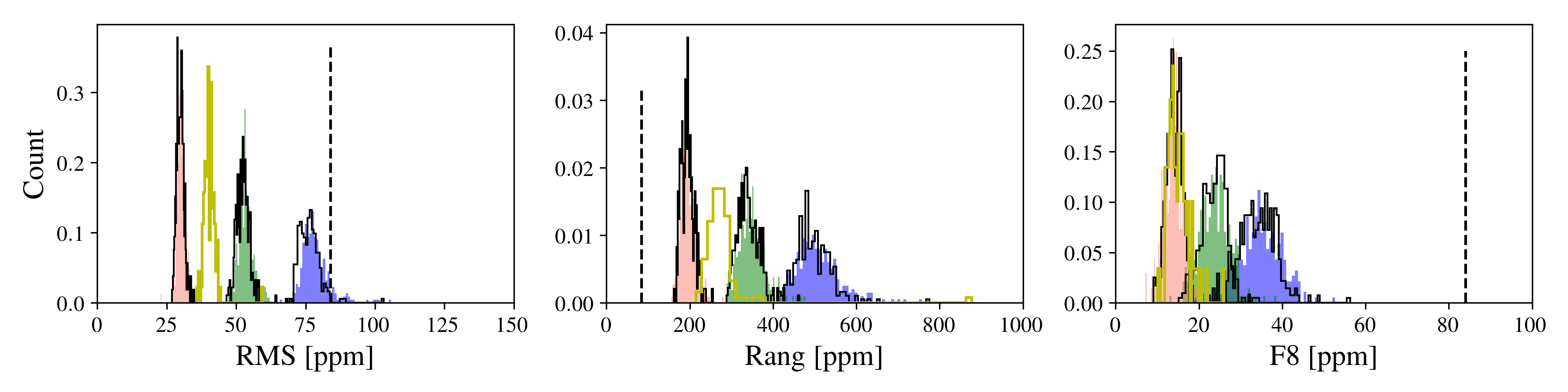}}
    \caption{
From left to right: Empirical distribution of the RMS, amplitude range, and F8 measurements evaluated on a set of one-day VIRGO subseries taken during a solar-cycle minimum (2008, colors are related to the three SPM channels). 
The empirical distributions obtained during a solar-cycle maximum (2003) are shown by the black contours. 
Distributions derived using the HMI data described in Sec.~\ref{Sec3} are shown by the yellow histograms.    
In each panel, the vertical dashed line indicates the transit depth expected for an Earth analogue ($84$ ppm). 
    }
    \label{Fig_distr}
\end{figure*}

By nature, the influence of a stochastic process cannot be known exactly. 
Consequently, it is only possible to make a probabilistic statement about its behavior.  
Here, to characterize the short-timescale variability in the time domain, we analyze  how its amplitudes are distributed. By ``amplitude'', here we mean the flux offset from the mean value of a given one-day subseries.

To do so, we first eliminated low-frequency noise by applying a Savitzky-Golay (SG) filter \citepads{doi:10.1021/ac60214a047}, parametrized by a window length of $15$ hours and a polynomial degree of $3$. This allowed us to consider only the frequency range $\nu<2 \times f_{\ell}$.
We note that, to avoid an inappropriate effect of the SG filter at the edges of the one-day subseries, we applied this low-frequency noise removal to the whole one-year time series (i.e., before dividing it into daily subseries as described in Sec.~\ref{Sec21}).
 This step is illustrated in Fig.~\ref{Fig_ts}, where we observe the wavelength dependence of these variations, which can reach several tens (red channel) to hundreds (blue channel) of parts per million \citepads{1988IAUS..132..239D}. 
The amplitudes of the flicker noise therefore increase when observing deeper layers of the solar photosphere \citepads{2005ApJ...623.1215J}.
Intuitively, we can expect this variability to follow a Gaussian distribution. Indeed, if we assume the dominant source of variability 
comes from convective motions, then this variability results from a  multitude ($\sim10^6$) of independent stochastic granules, each generating photometric 
variability of a similar amplitude. This is a physical manifestation of the central limit theorem. 
To validate this hypothesis, we performed a Shapiro-Wilk normality test 
\citepads{doi:10.1093/biomet/52.3-4.591} on each solar one-day subseries. This results in the nonrejection of the Gaussian hypothesis for more than $95\%$ of 
our dataset. Without establishing a formal proof of Gaussianity, this test does 
not argue against this hypothesis. 
In this approximation, we can therefore state that flicker noise is Gaussian (i.e., the amplitudes of the noise follow a Gaussian distribution). We 
note that Gaussianity here refers to the probability distribution 
of the amplitude of the flicker but does not inform on the way the power of this noise is distributed in the frequency domain (correlations). The latter remains, 
\textit{a priori}, unknown (see Sec.~\ref{Sec_24}). 
Gaussianity is important as it implies that noises correlated over timescales of 
minutes to hours (mostly convection processes) can be completely defined by the two first moments of their distribution: the mean and power spectral density (related to the variance, \citeads{Simon:2006:PDI:1212190}). 

Other common measurements to characterize a stochastic noise are the RMS and the amplitude range. 
For granulation, the `eight-hour flicker'' measure (or ``F8''), corresponding roughly to the RMS evaluated over a timescale between $0.5$ and $8$ hours, is also often computed.
To compute the F8 measurement as described in \citetads{2016ApJ...818...43B}, we binned the solar one-day subseries into intervals of  $30$ mins (to mimic the Kepler long-cadence observations) and used a $16$-point boxcar filter\footnote{We made use of the function ``convolution.Box1DKernel'' available from the Python package \url{http://www.astropy.org}.} to remove the long-term activity over periods $>8$ hr.
Figure~\ref{Fig_distr} shows the distribution of these three quantities evaluated 
on the one-day VIRGO subseries taken over a year. In each panel, the distributions obtained using subseries taken during a solar cycle minimum  (resp. solar cycle maximum) are shown by the colored (resp. black) histograms.
Typical RMS values are $30-40$ ppm (red), $50-60$ ppm (green), and $70-90$ ppm (blue). Extreme values can reach several hundred ppm ($200$, $350,$ and $500$ ppm for the red, green and blue channels, respectively). Although comparable to the global RMS, the F8 values are slightly smaller because they are based on data binned into  intervals of  $30$ mins and thus encapsulate less of the signal due to granulation. 
By comparing these values with the transit depth expected for an Earth analogue (see vertical lines), we gather that this variability can easily impede the transit detection or bias the inferred parameters. 

\begin{figure}[] \centering
    \resizebox{\hsize}{!}{\includegraphics{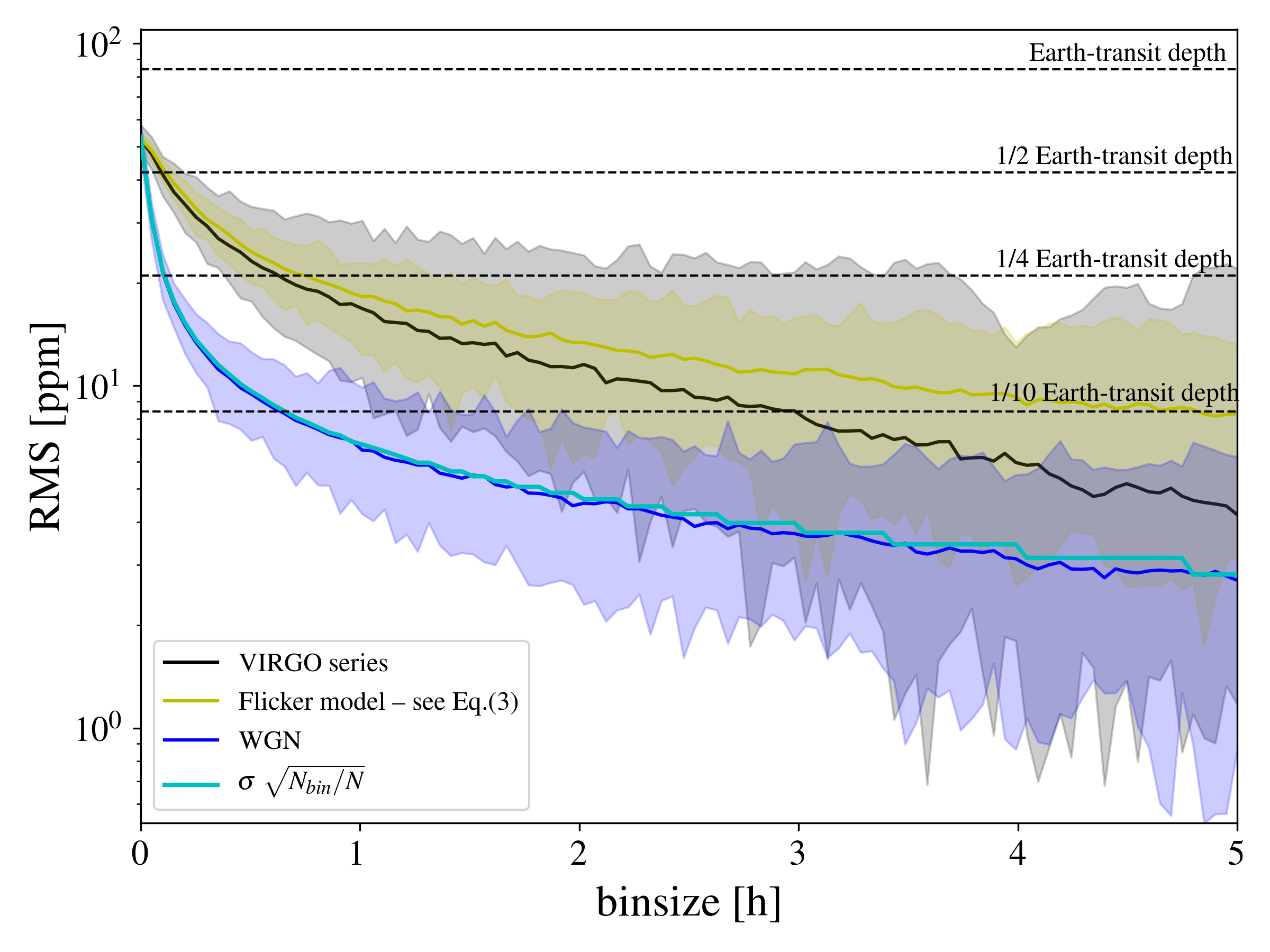}}
    \caption{Effect of data binning on the RMS of one-day solar subseries (1996, black), on the synthetic times series generated with Eq. \eqref{eq_flicker} (yellow), and on synthetic subseries of WGNs (blue). The RMS of the WGN series decreases as $\sqrt{N_{bin}/N}$ (light blue). Solid lines indicate the median values of the observed dispersion (shaded areas). The RMS of the time series at large bin sizes shows a significant dispersion as the number of data points decreases. Horizontal dotted lines indicate the fraction of a typical Earth-like transit depth (i.e., $84$ ppm).}
    \label{Fig_carac}
\end{figure}

In practice, binning the light curve of a single transit to decrease the impact of granulation noise is not optimal because of the long timescale of the stellar convection \citepads{2015A&A...583A.118M}. 
This is shown in Fig.~\ref{Fig_carac}, which illustrates the slower decrease of the RMS of the solar flicker noise compared to purely white noise. 
%where we observe that averaging the data only modestly decreases the amplitude of the flicker noise. 
%After a $30$-min data binning, the noise rms is decreased up to half of an Earth analogue's transit depth and only up to $1/10$ after a $3$-hrs data binning (see black curves). In comparison, the rms of a WGN with the same variance as the VIRGO observations will be decreased up to $1/4$ of an Earth transit depth with a $30$-min data binning and up to $1$ ppm after $3$-hrs binning (see blue curves).
For high-precision photometric data, the short-timescale variability cannot be approximated as a WGN. Instead, convection processes produce a colored stochastic noise. We note that acoustic modes are better described as deterministic noise sources (in the sense that they only affect specific frequencies depending on the stellar properties).

\begin{figure}[t] \centering
    \resizebox{\hsize}{!}{\includegraphics{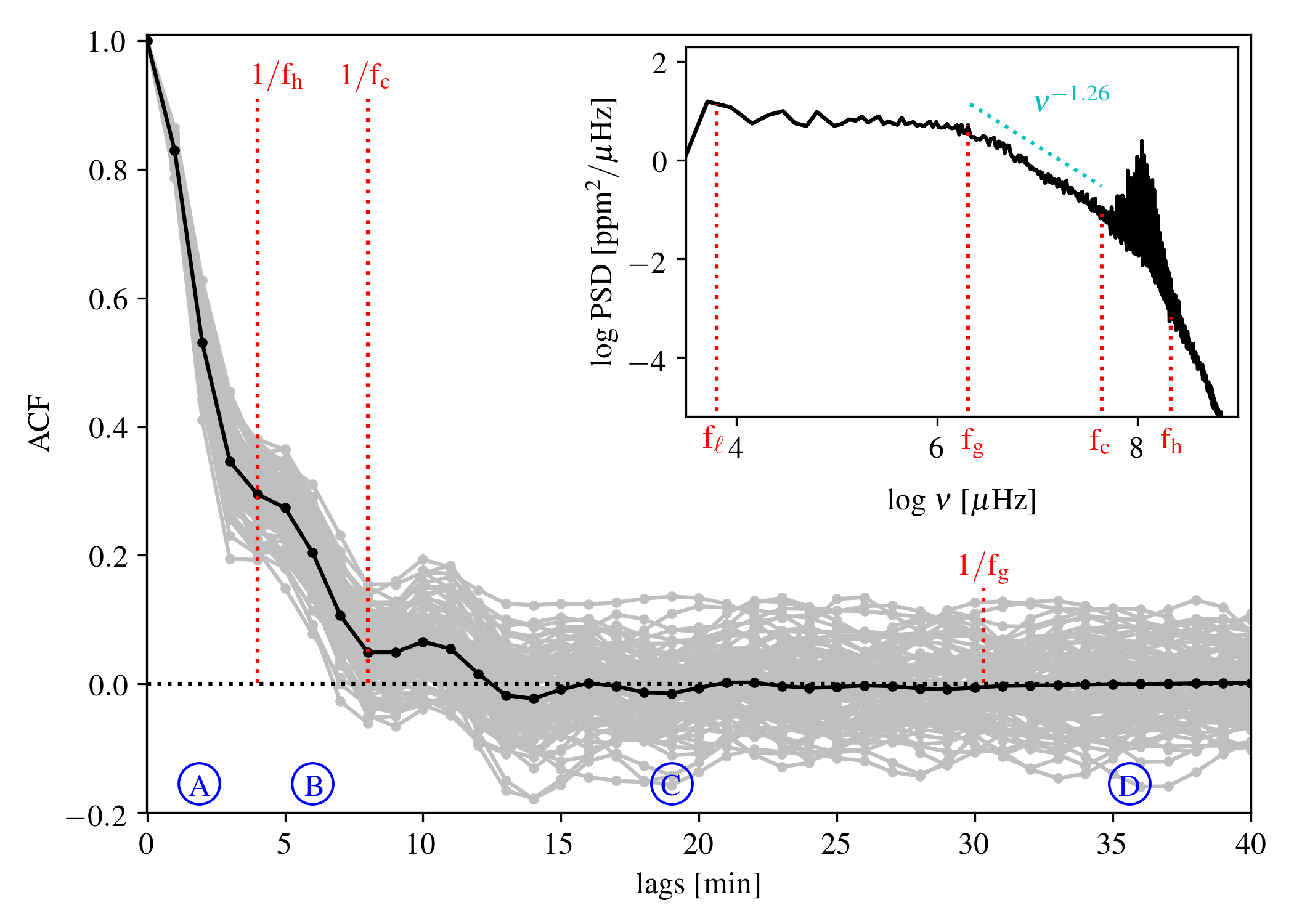}}
    \caption{Estimation of the splitting frequencies. Main panel: $L=364$ ACFs corresponding to the available solar one-day subseries observed in 1996 with the blue SPM channel (gray lines) and their mean ACF (black).
The mean ACF allows the automatic derivation of the upper and lower frequencies surrounding the acoustic mode regime $\{f_h,f_c\}$. 
Inset panel: Logarithm representation of the associated PSD estimates computed 
using Eq. \eqref{eq_PL}. The PSD allows us to localize the flicker frequency $\{f_g\}$. 
The lower frequency $\{f_\ell\}$ is set by the window length of the SG filter applied to each one-day subseries to remove the long-term variability.
In both panels, the splitting frequencies are indicated by the dotted vertical 
red lines. The blue dotted line represents the slope of the PSD (see Eq. \eqref{eq_DSP}) measured in the frequency region associated with the 
granulation regime $\nu\in [f_g,f_c]$. From high to low frequency, the distinct regions are denoted by letters A to D.}
    \label{Fig_acf}
\end{figure}

\begin{figure*}[b]
    \resizebox{\hsize}{!}{\includegraphics{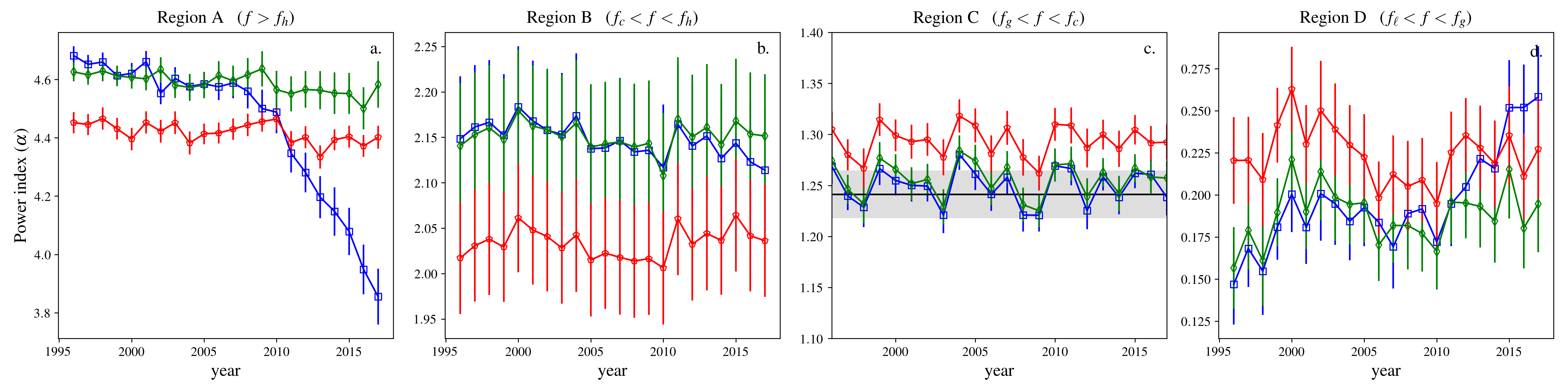}}
    \caption{Evolution of the inferred power index estimated on averaged 
periodograms of one-day solar subseries. From left to right: Index evaluated for 
(a) the high-frequency region, (b) the p-modes region, (c) the granulation region, and (d) the lowest frequency region.  Each color corresponds to an SPM channel. 
In region (c), the index obtained on the HMI solar observations discussed in Sec.~\ref{Sec31} is shown by the horizontal black line (with the $1\sigma$ uncertainties in gray). This index results from the averaged periodogram computed using the $91$ one-day subseries that have been selected during the period of a solar minimum (2017-2019).}
    \label{Fig_flicker}
\end{figure*}

\subsection{Frequency domain analysis: the solar power spectra}
\label{Sec_24}
% Check ACF/Ttot(lags) at lag zero give variance: Yes !
% Check Parceval's theorem: Ok!

The total dispersion of the time series is a scaled sum of the power at each frequency component (Parseval's identity, \citeads[see Eq. (6.1.7), p.169 of]{Li_2014}).
As discussed in Sec.~\ref{sec_22}, the solar PSD exhibits correlations over a multitude of timescales. For a given characteristic timescale, one noise process may dominate the others.\\

\noindent \textit{Acoustic mode timescales:} To localize the typical upper and lower frequency bounds of the acoustic modes, we use the autocorrelation function 
(ACF) of the observations, similarly to \citetads{Kallingere1500654}.
The first zero-crossing of the ACF is the autocorrelation time ($t_{c}$), that is, 
the time interval over which the noise de-correlates.
For the solar time series, $t_{c}$ corresponds to the \textit{corner frequency} 
($f_{c}$) and indicates the lower frequency of the p-modes (the separation between the regime dominated by the deterministic p-modes and those dominated by the stochastic convection). The upper limit of the p-mode signature ($f_h$) can be identified by the first dip in the ACF.
To precisely evaluate these frequencies, we computed the ACFs of each one-day subseries taken during one year and averaged them. 
The averaged ACF is shown in the main panel of Fig.~\ref{Fig_acf} (black), showing $f_h=4000$ $\mu$Hz ($\sim4$ min) and $f_c=2000$ $\mu$Hz ($\sim8$ min). \\

\noindent \textit{Convection timescales:}
The PSD of the VIRGO observations shows a ``knee'' shape in the frequency region $\nu<f_c$ (see inset panel of Fig.~\ref{Fig_acf}).  The physical origin of this kink is debated and is not observed in radial velocity measurements. Works based on Harvey-profile functions \citepads{1985ESASP.235..199H} that fit for the stellar background generally attribute this kink to the regime where supergranulation or mesogranulation\footnote{We note that the existence of intermediate scales of convective motions ($<3\times10^7$ m), known as mesogranulation, is  controversial even for the Sun \citepads{1989ApJ...342L..95S, 2000A&A...356.1050P}.} noise dominate. However, based on synthetic simulations of solar granulation alone, \citetads{2011A&A...532A.108S} attributed this frequency to the large dispersion of the granulation turnover timescales (the distribution of the granule lifetimes varies from $0$ to $30$ minutes).
We estimate this frequency (that we called the 
\textit{flicker frequency} in this work) at $f_g=555$ $\mu$Hz  ($30.3$ min). 
 We finally set the limit between the convection and the activity regimes (constrained here by the low-frequency SG filter with a width of $15$ hr) at $f_\ell=44.9$ $\mu$Hz  ($6.2$ h). \\
According to \citetads{2011A&A...532A.108S}, the physical properties of the stellar granule cells, such as the number of cells over the visible stellar surface, their photometric contrast ratio, their lifetime, and average size, can be extracted from the study of the observed stellar PSD. 
For example, we expect the frequency at which the PSD starts to become ``flat'' (i.e., $f_g$) to decrease with the increase of the granulation median turnover timescale. This would make this frequency a potential tracker for deriving the lifetime distribution of the granule cells over the stellar surface. Furthermore, the amplitude of the stellar PSD can be related to the average size of the granule cells. 
We should expect a decrease of the PSD amplitudes with the decrease of the size of the granules, as the  ratio between bright granules and intergranular lanes decreases \citepads{2011A&A...532A.108S}. 

In practice, the stellar PSDs are often modeled with Harvey functions \citepads{1985ESASP.235..199H}. As discussed in \citetads{10.1111/j.1365-2966.2012.20542.x}, the choice of this model is empiric and various slightly modified Harvey profiles are found in the literature (see e.g., \citeads{1985ESASP.235..199H,1999ASPC..173..297P,2004A&A...414.1139A,2014A&A...570A..41K, 2014ApJ...781..124C,2019MNRAS.489.5764P}). The difference between these models mainly comes from the choice of the power-law exponent, chosen to fit the observed stellar PSD \citepads{2011ApJ...741..119M}. Consequently, we chose here to represent the noise correlations by using a  simpler model based on $1/\nu^\alpha$ power laws \citepads{1999ASPC..173..297P}. We refer to this model as the flicker model in the following. Written in logarithmic scale, the model is:
\begin{equation}   
    \log P_L(\nu) = \alpha \log (\nu) + \beta,
    \label{eq_DSP}
\end{equation}
with $P_L$ being the yearly averaged periodogram of the one-day solar subseries defined 
as in Eq. \eqref{eq_PL}, $\alpha$ the power index, $\beta$ a constant, and $\nu\in \mathcal{R}_f$ the frequencies in the regions $\mathcal{R}_f :=\{ A, B, C, D\}$ as labeled in Fig.~\ref{Fig_carac}, corresponding to the different regimes outlined above.
The power index $\alpha$ and the constant offset $\beta$ have to be determined for each frequency region $\mathcal{R}_f$. 

We then estimated the parameter set $\{\alpha_i,\beta_i\}_{i=A,B,C,D}$ given in
Eq. \eqref{eq_DSP} through a least-square minimization of the solar PSDs. The average values are listed in Table.~\ref{Table1} and the estimated power indices for each of the $21$ years of VIRGO observations are shown in Fig.~\ref{Fig_flicker}. 
\begin{table*}[t]
\centering
\caption{Average values of parameters $\{\alpha, \beta\}$ corresponding to Eq.  
\eqref{eq_DSP} for regions A, B, C, and D and for observations in the three SPM 
channels. }
\begin{tabular}{|c|c|c|c|c|c|c|c|}
\hline
Region & Frequency & \multicolumn{3}{c|}{$<\alpha>~\pm~\Delta \alpha$} & 
\multicolumn{3}{c|}{$<\beta>~\pm~\Delta \beta$} \\ \cline{3-8}
& interval       & Blue  &  Green   & Red    & Blue   & Green    & Red\\
 \hline
A&$\nu>f_h$       &$4.44 \pm 0.24$ & $4.59 \pm 0.03$ & $4.42 \pm 0.03$ & $-27.18 \pm 1.37$ & $-28.76 \pm 0.21$ & $-28.87 \pm 0.18$\\
B&$f_h>\nu>f_c$     &$2.15 \pm 0.02$ & $2.15 \pm 0.01$ & $2.03 \pm 0.02$ & $-14.34 \pm 0.15$ & $-15.15 \pm 0.11$ & $-15.61 \pm 0.11$\\
C&$f_c>\nu>f_g$    &$1.25 \pm 0.02$ & $1.26 \pm 0.02$ & $1.29 \pm 0.01$ & $-8.68 \pm 0.12$ & $-9.50 \pm 0.11$ & $-10.94 \pm 0.10$\\
D&$f_g>\nu>f_\ell$ &$0.20 \pm 0.03$ & $0.19 \pm 0.02$ & $0.22 \pm 0.02$ & $-0.81 \pm 0.22$ & $-1.51 \pm 0.14$ & $-2.91 \pm 0.13$\\
\hline
\end{tabular}
\label{Table1}
\end{table*}
As visible in the PSD, the highest values of the power index $\alpha$ are found 
for the highest frequency region (A).
For all regions considered, the power indices observed in the red channel appear relatively constant with time. For the green and blue 
channels however, which are more strongly influenced by the detector ageing, the power index drops significantly in recent years ($>2008$) in the high-frequency regions A and B. We note that this effect is likely purely instrumental due to increasing white noise. 
For the timescales inferior to several hours (regions A, B and C), the index values do not vary along the solar cycle but they do for region D. Once again, these results illustrate both the global high quality of VIRGO observations and the stationarity of the flicker variability. Moreover, for $\nu < f_c$ (regions C and D), the similarities between the time series observed in different colors indicate similar correlations independently of wavelength. 
Consequently, alternatively to Harvey models, a flicker-based model can be designed based on simple power laws as in Eq. \eqref{eq_DSP}. Synthetic time series of this stationary stochastic process can then be generated as:
\begin{equation}
    I(t) = \sum_{k=0}^{\infty} \sqrt{2~S(\nu_k)~\delta\nu} ~ \cos{(2\pi \nu_k t 
+ \phi_k )},
    \label{eq_flicker}
\end{equation}
with $S(\nu) = \exp(\beta/\nu^\alpha)$ being the parametric noise PSD based on 
Eq. \eqref{eq_DSP}, $t$ the time of the observations, $\delta\nu$ the sampling 
frequency, and $\phi$ the random phase $\in [0,2\pi]$. Following 
Eq. \eqref{eq_flicker}, we generated synthetic time series based on parameters
listed in Table.~\ref{Table1}. We compared these synthetic series with solar observations in Fig.~\ref{Fig_carac}. As soon as the number of data points is sufficiently large (i.e., at bin sizes $<3$ hr), we observe a similar behavior for both series demonstrating that our simple flicker-noise model is realistic enough to provide a first-order approximation of short-timescale solar variability.
%

%%%%%%%%%%%%%%%%%%%%%%%%%%%%%%%%%%%%%%%%
% SEC.III: EMPIRICAL (OPTIMAL) BIAIS ON TRANSIT DEPTH
%%%%%%%%%%%%%%%%%%%%%%%%%%%%%%%%%%%%%%%%

\section{Impact of solar short-timescale variability on inferred transit parameters}
\label{Sec3}

Before extracting the statistical properties of flicker noise from stars other than the Sun, we aim to evaluate its impact on the parameters inferred from exoplanet transit light curves. 
To this end, we generated artificial light curves of planetary transits based on resolved images of the solar disk during a solar-cycle minimum. In this section, we describe the analysis of these light curves assuming purely white Gaussian noise and quantify the errors made on the inferred parameters when the correlation properties of flicker noise are not correctly taken into account.

\subsection{Artificial transit light curves in HMI observations}
\label{Sec31}

In order to generate realistic planetary transit light curves, we selected observations\footnote{\url{http://jsoc.stanford.edu/ajax/exportdata.html}} 
taken by the \textit{Helioseismic and Magnetic Imager} (HMI) instrument onboard the \textit{Solar Dynamics Observatory} (SDO).
Since 2010, HMI has been observing the photospheric Fe I absorption line at $617.3$ nm almost continuously, producing one image of the solar disk every $45$ seconds. 
The resolution of HMI is 0.505 arcsec/pixel and the optical resolution is 0.91 arcsec, corresponding roughly to 366 km on the solar surface at disk center \citepads{2012SoPh..275..229S}.

We selected $91$ different dates during a minimum of the solar activity cycle where no significant signatures of active structures (spots, plages) were observed on the visible part of the solar surface. For each date, we downloaded images spanning one day and extracted the solar flux from each of them by integrating the intensity over all pixels. 

\begin{figure} \centering
\resizebox{\hsize}{!}{\includegraphics{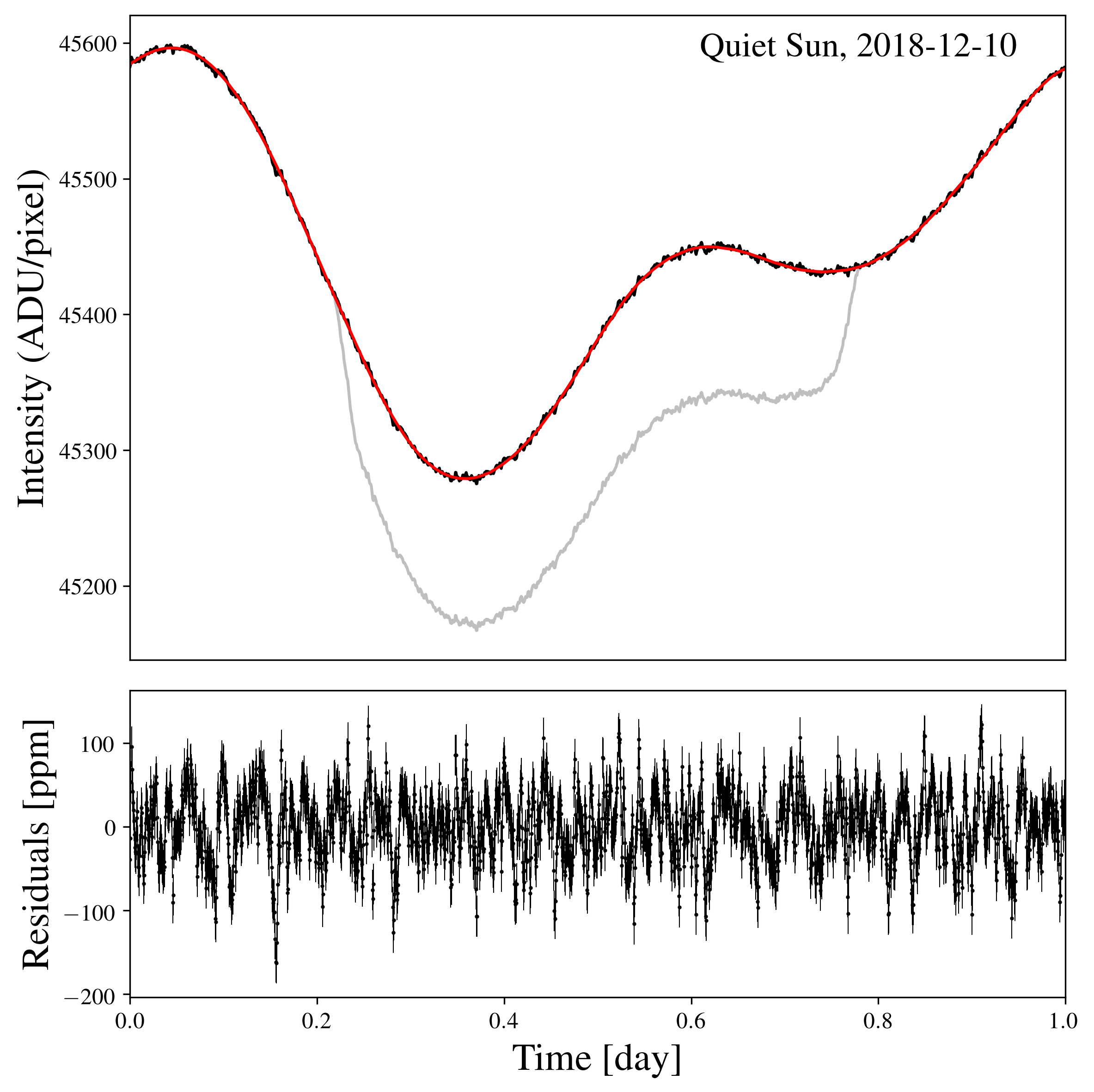}}   
\caption{Top: Example of a raw solar light curve extracted from HMI images (black) and the corresponding smooth detrending function based on SG filters with a window size of $5$ hours (red). An artificial transit of a $5~R_\oplus$ planet crossing the center of the solar disk ($b=0$) is shown in gray. Bottom: Residuals of the raw solar light curve after correcting by the SDO satellite motion.}
    \label{Fig_SDO_orbit}
\end{figure}

To create artificial transit light curves of exoplanets in each one-day solar dataset, we superimposed a black sphere on the images, and moved it across the disk, mimicking a transiting planet. We assume exoplanet sizes of $R_p=1$, $3$, $5$, $7$, and $10$ Earth radii ($R_\oplus$). 
For each planet size, we simulated transits with impact parameters $b = 0$, $0.2$, $0.4$, $0.6$ and $0.8$. 
This led to $25$ artificial transit light curves for each of the $91$ solar time series. 
One of these, corresponding to $R_p = 5$ $R_\oplus$ and $b = 0,$ is shown in the top panel of Fig.~\ref{Fig_SDO_orbit} (see gray line).

This way of generating artificial transit light curves has three effects that are extraneous to real exoplanet transits.
First, since the SDO spacecraft is on an inclined geosynchronous orbit around Earth, the apparent size of the Sun on the CCD images changes with time. 
Second, as the size of the planet is scaled in terms of pixels, the planet-to-sun radius ratio ($p=R_p/R_\odot$) slightly changes over time.
To increase the accuracy of $p$, which is  needed to assume that the radius ratio is constant over time, we oversampled the artificial exoplanet by a factor of five. This means that on one pixel of the HMI image there are $5\times5$ pixels in the exoplanet mask\footnote{We note that with this oversampling, a synthetic Earth-like planet disk corresponds to $23200$ (oversampled) HMI pixels (compared to $928$ without oversampling). For comparison, a typical convection cell size ($1$Mm) is roughly $13$ times smaller than the Earth diameter ($12.7$ Mm).}. The exoplanet mask was then derived in the high-resolution oversampled regime, and interpolated back to the original image resolution under conservation of flux within each pixel. 
This leads to partial pixel eclipses at the boundary of the exoplanet, and consequently to a greatly enhanced accuracy. 
In Appendix~\ref{AppA}, we show the intrinsic variation of the transit depth ($p^2$) over time, which is $<0.02\%$ of the true value, leading to relative errors between the transit models and the transits resulting from our experiment below $2$ ppm.
Finally, with this experiment, we assume discrete exoplanet positions for each HMI image, neglecting the movement of the planet during HMI's $45$s exposures. While the effect of exposure time is known to affect the transit parameters \citepads{2010MNRAS.408.1758K}, we assume our temporal cadence to be sufficiently small to do not impact our resulting light curves.

\begin{figure*}[t]
    \resizebox{\hsize}{!}{\includegraphics{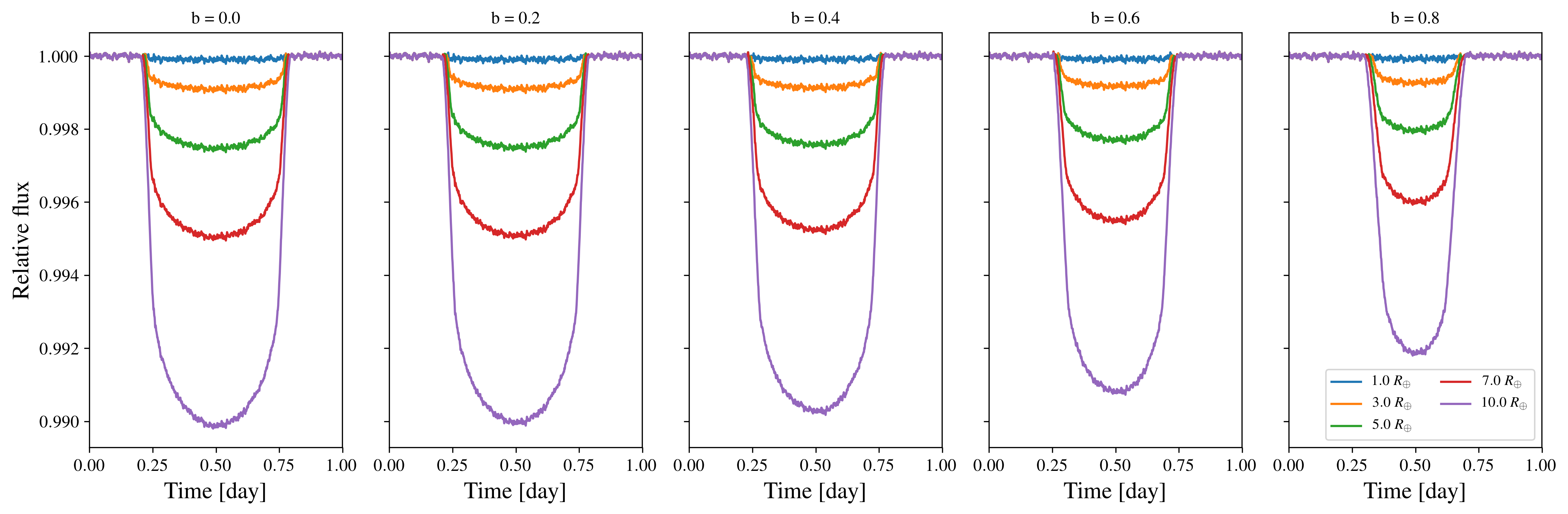}}
    \caption{Example of artificial exoplanet transit light curves generated using solar HMI observations (quiet Sun, 2018-12-10). Each panel shows five transits for planets with sizes ranging from $1$ to $10$ $R_\oplus$ (see legend). Panels from left to right display different orbit configurations (see the panels' header).}
    \label{Fig_transits}
\end{figure*}

\begin{figure}[t] \centering
    \resizebox{\hsize}{!}{\includegraphics[width=\linewidth]{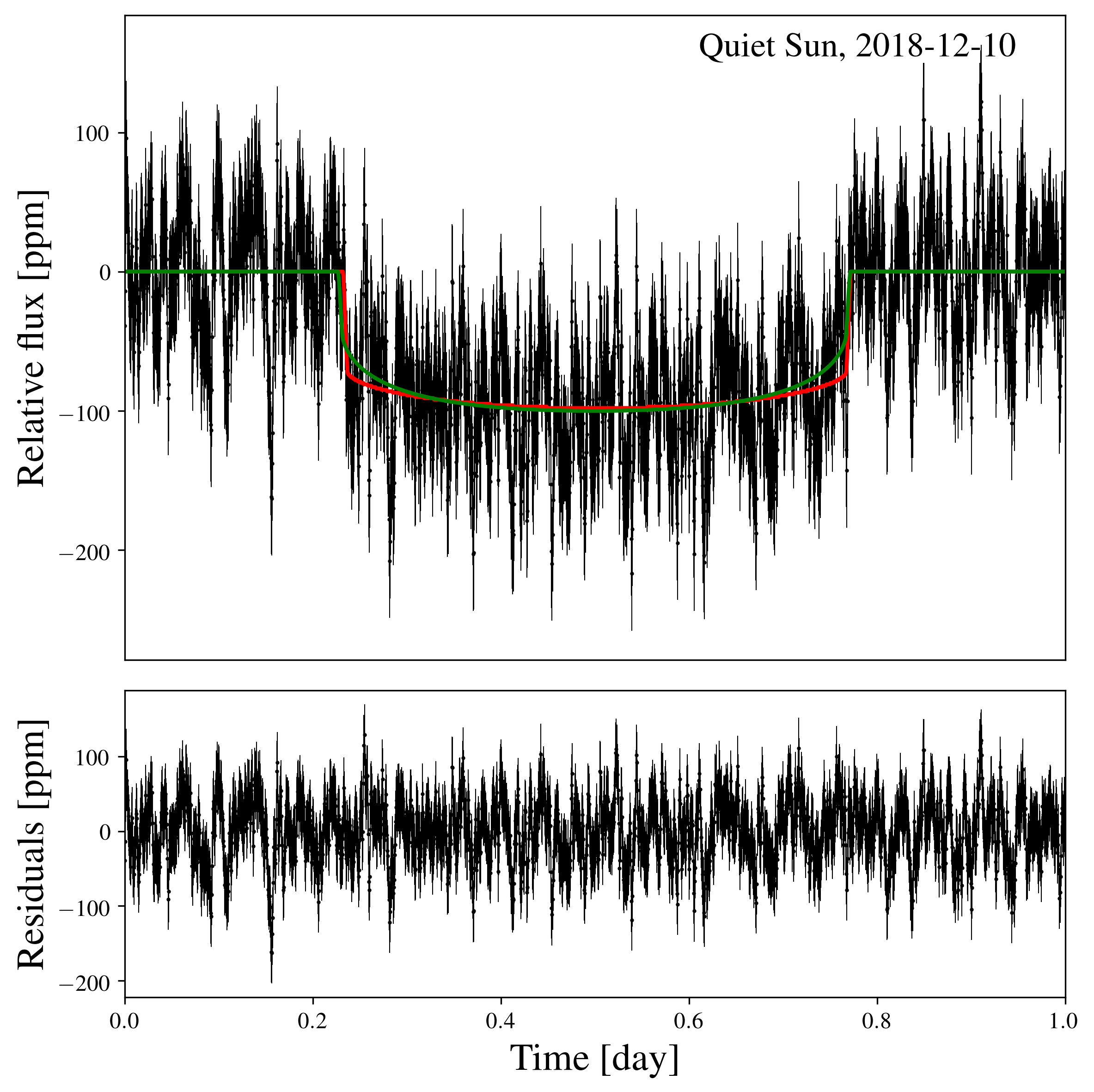}}
    \caption{Top: Example of artificial transit of an Earth-sized planet crossing the disk center of the Sun ($b=0$, black). The transit model with the true input parameters is shown in green and the model computed using the inferred parameters in red. The error on  $R_p/R_s$ is around $2\%$ in this example. Bottom: Residuals based on the inferred transit model.}
    \label{Fig_transits_Earth}
\end{figure}

The effect induced by the orbit of the recording satellite (period $\sim$ one day) is visible in Fig.~\ref{Fig_SDO_orbit} as a long-term quasi-sinusoidal variation.
From the analysis based on VIRGO data (see Sec.~\ref{Sec2}), we know that granulation noise dominates periods of $<30$ min. 
To correct the light curves from the effect of the satellite motion, we chose to filter each raw (i.e., without transits) solar time series with a smooth SG filter that has a passband larger than  ten times the characteristic period of granulation (i.e., $>5$ hours). 
An example of the final residuals resulting from this data filtering is shown in the bottom panel of Fig.~\ref{Fig_SDO_orbit}. 
The corrected transit light curves are shown in Fig.~\ref{Fig_transits} (see also Fig.~\ref{Fig_transits_Earth} for a better display of a central transit of an Earth-sized planet).
We note that we also tried to perform this correction using GP, but to avoid any influence of the GP on the flicker noise, we chose to use simple smooth functions.

Finally, the HMI observations are given without error bars, which are necessary to derive the transit parameters (see Sec.~\ref{Sec32}) together with 
their uncertainties. 
We estimated the errors on the individual data points from the residual scatter of the transit-free light curves after correction of the variation due to the satellite motion. On average, we obtained $\sigma=20-30$ ppm (see Appendix.~\ref{AppB}). 
The whole set of artificial transit light curves is publicly available at \url{https://doi.org/10.5281/zenodo.3686871}.

%================================================================
\subsection{Impact of flicker on the inferred transit parameters}
\label{Sec32}
%================================================================

To retrieve the transit parameters, we used state-of-the-art transit modeling based on the \citetads{2002ApJ...580L.171M} algorithm. 
For each artificial transit light curve associated to a given set of solar observations, we performed a Markov Chain Monte Carlo (MCMC) analysis following the scheme described in \citetads{2017A&A...606A..18L} and \citetads{2020MNRAS.492.1761L} and using the differential-evolution MCMC engine developed by \citetads{2017AJ....153....3C}.

We assumed the planet orbital period ($1$ year) was known and fitted for the planet-to-star radius ratio ($R_p/R_s$), the epoch of mid-transit ($T_0$), the impact parameter ($b$), the transit duration ($t_d$), and the quadratic limb darkening coefficients ($u_1$, $u_2$). We applied uniform priors to each of these parameters. The results are discussed in the following. 

\begin{figure*}[pt]\centering
    \resizebox{\hsize}{!}{\includegraphics{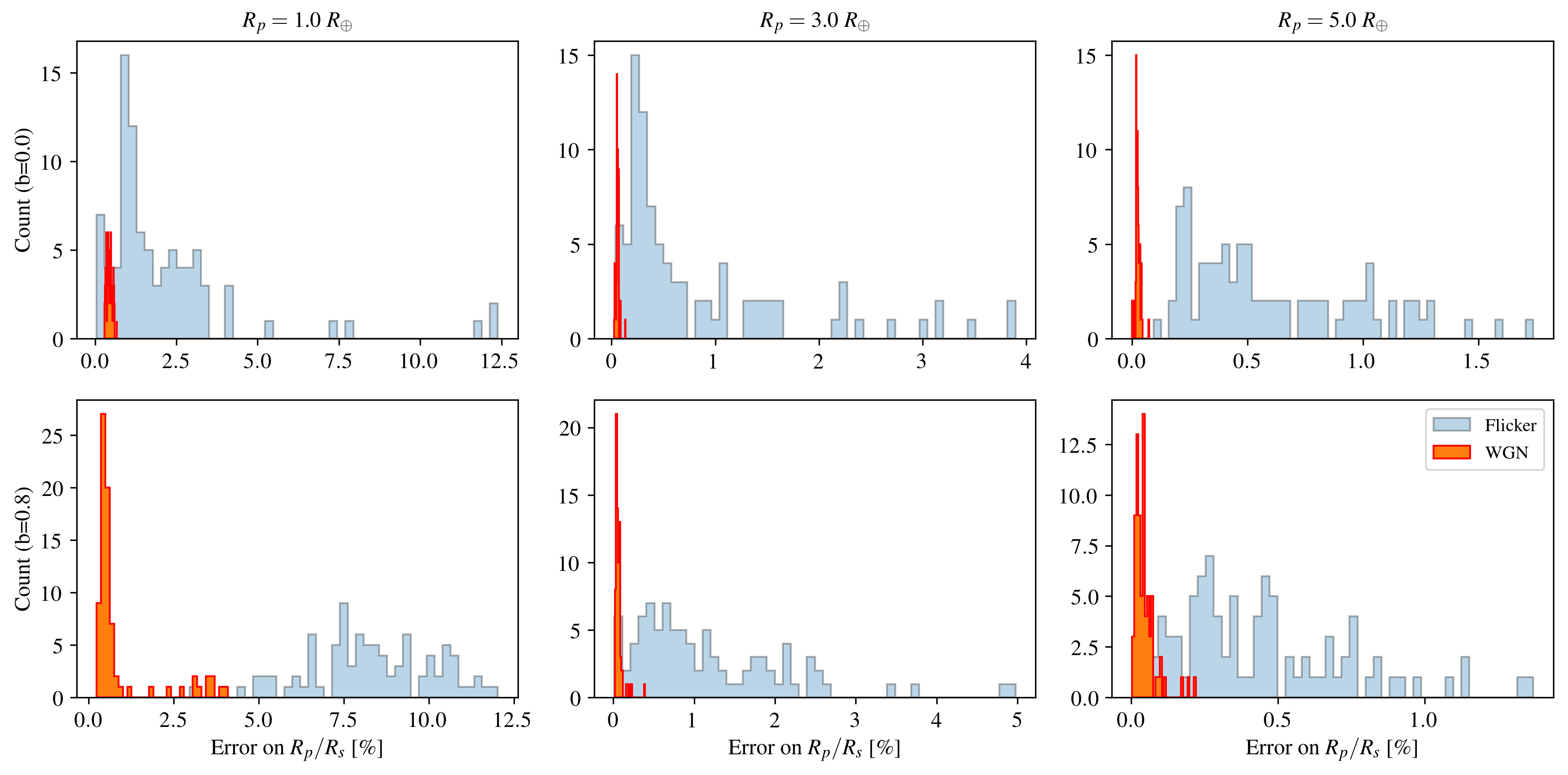}}
    \caption{Distribution of the absolute percentage error on the planet-to-star radius ratio inferred from our simulated transits for planets with sizes $R_p=[1,3,5]$ $R_\oplus$ (left to right) and impact parameters $b=0$ (top) and $b=0.8$ (bottom). Inferred errors on radius ratios derived from light curves containing only WGN are shown in red. }
    \label{Fig_r1}
\end{figure*}

We found relative errors on $R_p/R_s$ increasing with decreasing planet size, which is expected as larger planets create deeper transits while the noise level remains similar. Absolute percentage errors (difference between the peak of the MCMC posterior and the true value normalized by the true value) on $R_p/R_s$ appear small ($<1$ -- $2\%$ for planets with sizes above $3~R_\oplus$), but can be large ($\sim10\%$) for Earth-sized planets (see Fig.~\ref{Fig_r1}). 
For comparison, we generated synthetic light curves containing only WGN and derived the errors on $R_p/R_s$ using the same MCMC approach. To generate these synthetic WGN time series, we isolated the high-frequency noise present in the data by (i) removing the true transit model from our time series, (ii) applying a SG filter to filter out correlated noise at timescales above $15$ min, and (iii) generating a transit light curve with the \citetads{2002ApJ...580L.171M} analytical model.
For all involved transit parameters, we found errors on $R_p/R_s << 1\%$ (see red histograms in Fig.~\ref{Fig_r1}).  

\begin{figure*}\centering
\resizebox{0.9\hsize}{!}{\includegraphics[width=\linewidth]{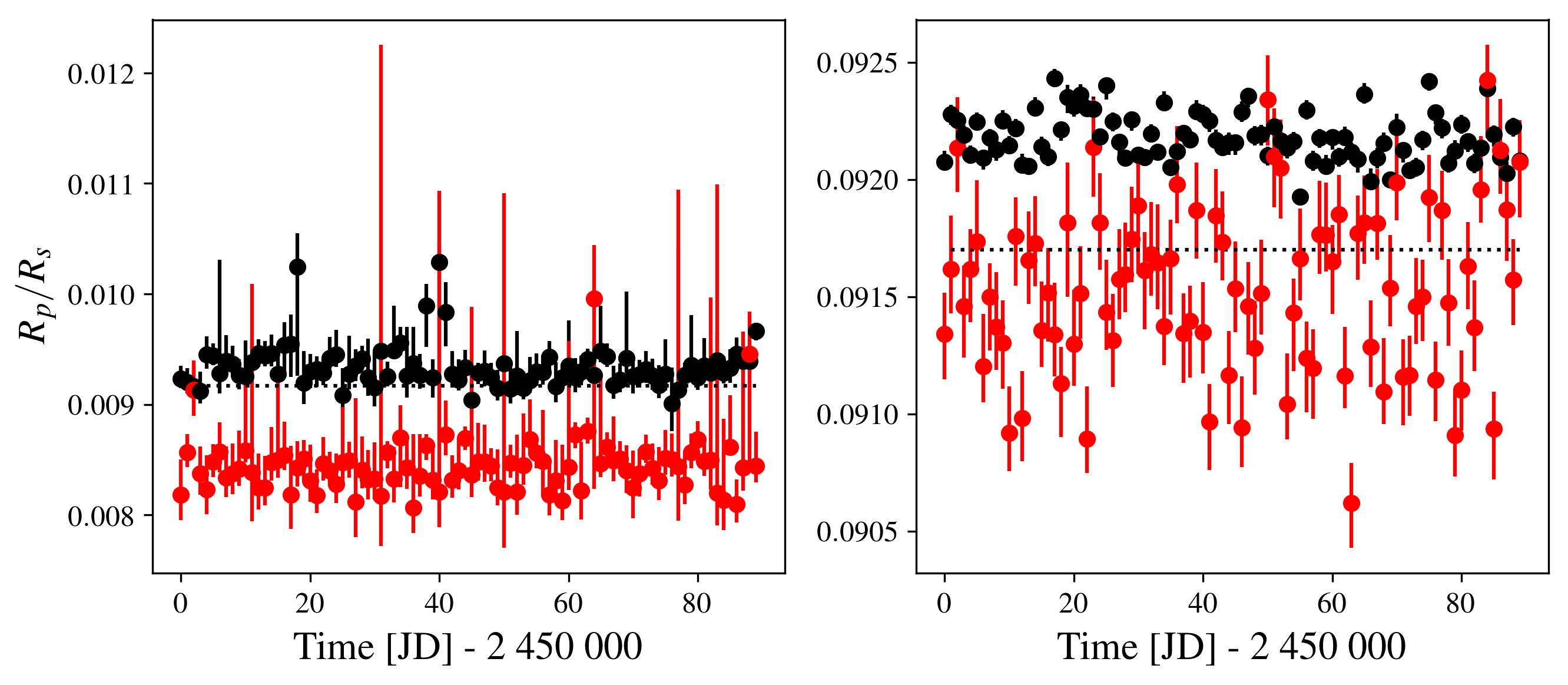}} \\
\caption{Planet-to-star radius ratio and $1\sigma$ uncertainties inferred from the MCMC analyses performed on the artificial light curves of exoplanets of size $R_p=1$ $R_\oplus$ (left) and $R_p=10$ $R_\oplus$ (right). The case of $b=0$ is shown in black and $b=0.8$ in red. The true radius ratios are indicated by the horizontal dotted lines. }
    \label{Fig_results_global2}
\end{figure*}

We now focus on the impact of the solar flicker noise on the inferred planet-to-star radius ratios. Figure~\ref{Fig_results_global2} (left panel) shows that for Earth-sized planets, the true values are not within  the $1\sigma$ uncertainties for $49\%$ of the cases when $b=0$ and for $81\%$ when $b=0.8$. 
Moreover, the true  value is not even included within the $3\sigma$ uncertainties for $11\%$ of the realizations when $b=0$ and for $4.5\%$ when $b=0.8$. 
For the largest exoplanets ($10$ $R_\oplus$, see right panel), while the percentage error is very small ($<1$ $\%$), none of our inferred values 
contain the true radius ratio within their $1$ and $3\sigma$ uncertainties when $b=0$ . This offset is not observed when the planet crosses the 
solar limb ($b=0.8$). This effect is likely due to uncertainties on the limb darkening parameters that can bias the retrieved transit parameters \citepads{2016MNRAS.457.3573E}. 
In line with this result, \citetads{2017AJ....153....3C} showed that MCMC analyses that ignore time-correlated noise produce inaccurate transit-depth estimates and can largely underestimate their uncertainties. Both effects increase as the variance of the correlated noise increases.

\begin{figure*}[hbtp!]\centering
\resizebox{\hsize}{!}{\includegraphics[width=\linewidth]{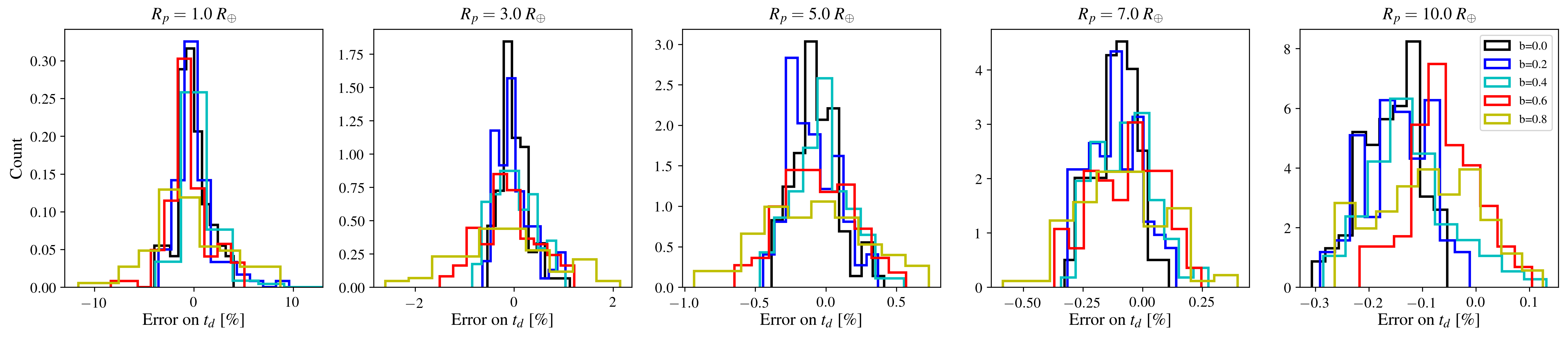}} \\  
\resizebox{\hsize}{!}{\includegraphics[width=\linewidth]{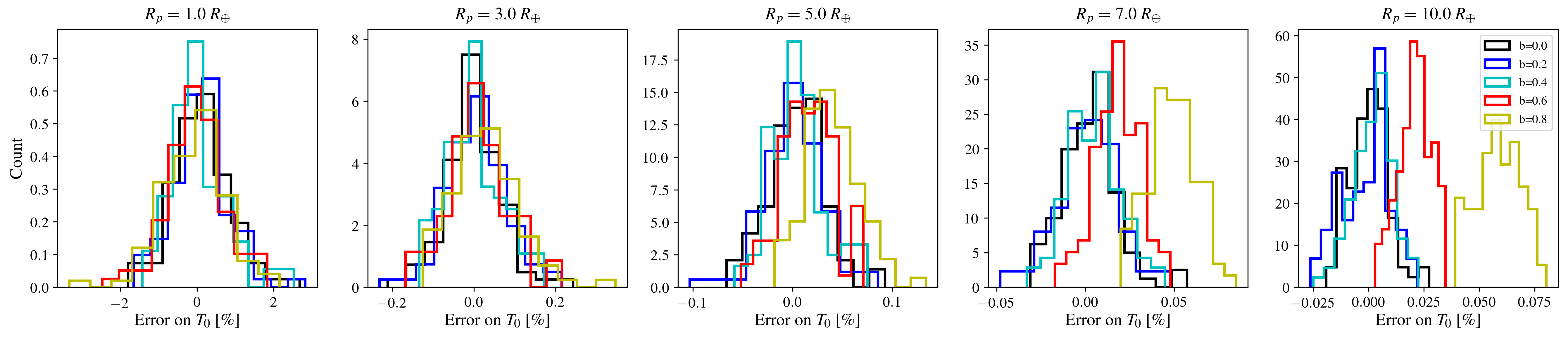}} \\
\resizebox{\hsize}{!}{\includegraphics[width=\linewidth]{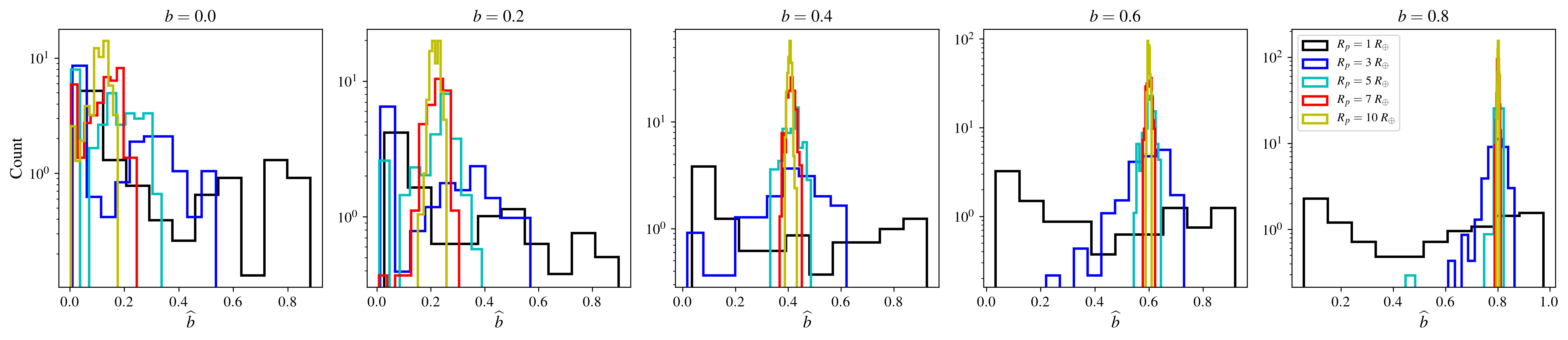}} \\  
\caption{From top to bottom: Distribution of the absolute percentage errors on $t_d$, $T_0$, and $b$ for the whole set of artificial transit light curves.}
\label{Fig_results_global}
\end{figure*}

For the transit duration, we found a distribution of the relative errors centered around $0\%$, and with a dispersion up to $2\%$ for the Earth-sized planet (see first row of Fig.~\ref{Fig_results_global}). For this planet, we measure a dispersion around the true value ($t_d=13.09$ hr) of $\pm 31.3$ min for $b=0$ (i.e., error up to $4\%$) and of $\pm 55.4$ min (i.e., error up to $10\%$) for $b=0.8$ (with the true value $t_d=7.97$ hr).

For the time of mid-transit, $T_0$, we found no offset but a dispersion of the inferred parameters around the true value that is also quite large for the Earth-like planet: $\pm 19.2$ min for $b=0$ and $\pm 15.4$ min for $b=0.8$ (see second row of Fig.~\ref{Fig_results_global}).  Moreover (not shown here), the true $T_0$ was not contained inside the $1\sigma$ uncertainties for more than $76\%$ of the cases when $b=0$ ($28\%$ fell  outside the $3\sigma$ uncertainties) and $77\%$ for $b=0.8$ ($25\%$ fell outside the $3\sigma$ uncertainties). These errors can directly affect the measurement of transit timing variations in multi-planet systems.

Finally, for the impact parameter, the difference between the true and inferred value decreases with the size of the planet (see last row of Fig.~\ref{Fig_results_global}). For the Earth-sized planet, the distribution is almost uniform making the inferred impact parameter essentially unconstrained by the observations.

These results suggest that previous analyses of single transit events performed on light curves dominated by stellar noise (e.g., for bright solar-like Kepler targets, see \citeads{2011ApJS..197....6G,2015AJ....150..133G}) may be miscalculated by a few percent due to a misunderstanding of the correlation structure of the short-timescale stellar noise. 
All these errors can directly impact the characterization of the exoplanet. As a consequence, they can impact models of the interior structure of planets (see e.g., \citeads{2015A&A...577A..83D}) and atmosphere. 
The present experiment demonstrates the need for new statistical tools dedicated to correctly accounting for the effect of flicker noise, and strategies to mitigate its effect on the derived transit parameters. These new tools will need to encapsulate the statistical properties of flicker that have been derived in this paper (see Sec.~\ref{Sec2} for the Sun and the Sec.~\ref{Sec4} for other main sequence stars). 

Some studies have already focused on identifying the component of flicker noise in photometric data and modeling it. Recently, \citetads{2020arXiv200208072M} performed a similar study, also using HMI data to reproduce exoplanet transits. This study conceptually differs from ours: they used a single HMI image to estimate the flicker variability amplitudes. As a consequence, they underestimate the flicker noise within the resulting light curves. Then, to probe the effect of short-timescale stellar variability on transits, they injected model transit light curves into transit free data. The strength of our study lies in the fact that we extract the effect of this variability on planetary transits without any prior assumptions (e.g., limb darkening) of the transit light curve shape, and at the same time we  also account for variability introduced by the changing photospheric flux obscured by the planet.

Using 3D radiative hydrodynamical simulations of solar convection,  \citetads{2015A&A...576A..13C} succeeded to evaluate the contribution of flicker on the transit light curve of Venus 
(which appeared in 2004). Moreover, these 3D simulations allow us to derive very accurate limb-darkening laws \citepads{2015A&A...576A..13C}, which is another critical aspect we must 
consider for deriving unbiased transit parameters as described above. 

Other studies proposed to use GP to model the short-timescale variability noise sources (including flicker, oscillations p-modes, and high-frequency noise; see e.g., \citeads{2015ApJ...800...46B,2019MNRAS.489.5764P}). Compared to analyses based on WGN models (as in this section), modeling the flicker noise with GP models increases the parameter uncertainties but could allow us to improve the accuracy on the transit parameters. However, this modeling approach alone does not improve the precision of the transit parameters. 
A noteworthy advantage of the GP approach lies in its capacity to constrain the stellar noise properties. For example, \citetads{2019MNRAS.489.5764P} found that their GP regression is able to derive accurate values of the p-mode mean oscillation frequency $\nu_{max}$. 
Following this same idea of linking the properties of the flicker noise with the stellar parameters, we now investigate the correlations associated to flicker using Kepler observations for a range of very bright stars on (or near) the main sequence.

%%%%%%%%%%%%%%%%%%%%%%%%%%%%%%%%%%%%%%%%%
% SEC.IV: FLICKER ANALYSIS (KEPLER DATA)
%%%%%%%%%%%%%%%%%%%%%%%%%%%%%%%%%%%%%%%%%

\section{Short-timescale stellar variability on Kepler stars}
\label{Sec4}

In this section, we aim to extract the granulation properties using Kepler observations of bright stars. 
The flicker amplitude is already known to be related to the stellar parameters (see e.g., \citeads{2016ApJ...818...43B}, based on F8 measurements).
Therefore, we focus here on the relation between the flicker characteristics timescales and the stellar fundamental parameters.  

\subsection{Kepler short-cadence observations}
\label{Sec41}

The Kepler prime mission was operating from 2009 to 2013 \citepads{2010Sci...327..977B}. It operated in the optical wavelength range $\lambda \in [400,865]$ nm: a much broader passband than that of the solar observatories VIRGO and HMI (see Sec.~\ref{Sec2} and \ref{Sec3}). To compare Kepler images with solar observations from VIRGO, previous studies often used a sum of the red and green channels as these are the closest to the Kepler passband (see e.g., \citeads{2013ApJ...769...37B, 
2017A&A...608A..87S}). 
Kepler long-cadence observations ($29.4$ min) have been intensively studied to derive the long-timescale stellar variability for stars of different stellar types \citepads{2013ApJ...769...37B, 2012A&A...539A.137M} as well as the short-timescale variability evolving periods of less than one day (see e.g., \citeads{2011ApJ...741..119M};  \citeads{2014ApJ...781..124C}; \citeads{2016ApJ...818...43B}; \citeads{2018MNRAS.480..467P}).
For the latter, it has been shown that the granulation amplitude (F8 measurements) can be linked to the stellar surface gravity. The same can be applied for the turnover granulation timescales (i.e., the period corresponding to $f_c$, \citeads{Kallingere1500654}).

In this section, we focus on the short-cadence (SC, $58.8$~s) Kepler observations that have been performed on a small number of stars,
as long-cadence observations do not carry information about noise correlations at timescales below $30$~min, where most of the granulation signal is located. 
We focus here on the determination of the correlation properties that we defined in Sec.~\ref{Sec_24} through a flicker power index $\alpha_g$ (measured as the slope of the  PSD in the frequency range between the corner and flicker frequencies, $f_c$ and $f_g$, respectively).

For this purpose, we selected the brightest stars observed by Kepler in SC mode (apparent magnitude $m_v<11.5$), for which no planet has been detected (a total of $3970$ objects).
From this sample, we removed binary stars, rotationally variable stars, stars in clusters, peculiar stars, and red giants (ending up with $1401$ objects). At this point, our sample contained mainly G and F stars as late-type stars (K to M) were rejected by the magnitude cutoff.
For each star from our sample, we downloaded the whole set of SC observations and detrended the light curves following a similar procedure as for the VIRGO time series (see Sec.~\ref{Sec2}).
As the Kepler spacecraft rotated by $90$ degrees every $90$ days (to keep the solar panels in the direction of the Sun), the observations of a given star are divided into four subseries (called ``quarters'') per year. 
For each target and quarter, we carry out the following procedure: (1) We remove the data points affected by spacecraft safe-mode events and corrected for the background; (2) we smooth the time series using a running average of $3$ days length and localize the $3\sigma$ outliers; (3) excluding these outliers, we bin the resulting time series into intervals of  $24$ hr, apply a spline function to this binned series, and use this function to normalize the initial time series (containing the outliers of step 2.);  and (4) finally we remove the $5\sigma$ outliers from the time series.

Following this procedure we combined all observations of the same target from all quarters and split the final detrended time series into one-day subseries. For each one-day subseries, we filtered out the low-frequency noise associated with magnetic activity using a SG filter with a $15$ hr passband.  

For all targets, the photometric contribution of granulation to the high frequencies (HF) cannot be observed (contrary to VIRGO observations; see region $\nu < 5000~\mu$Hz in Fig.~\ref{Fig_periodo}).
Indeed, the HF noise in these Kepler data is non-negligible with an amplitude comparable to granulation noise.

\begin{figure}[t] \centering
    \resizebox{\hsize}{!}{\includegraphics[width=\linewidth]{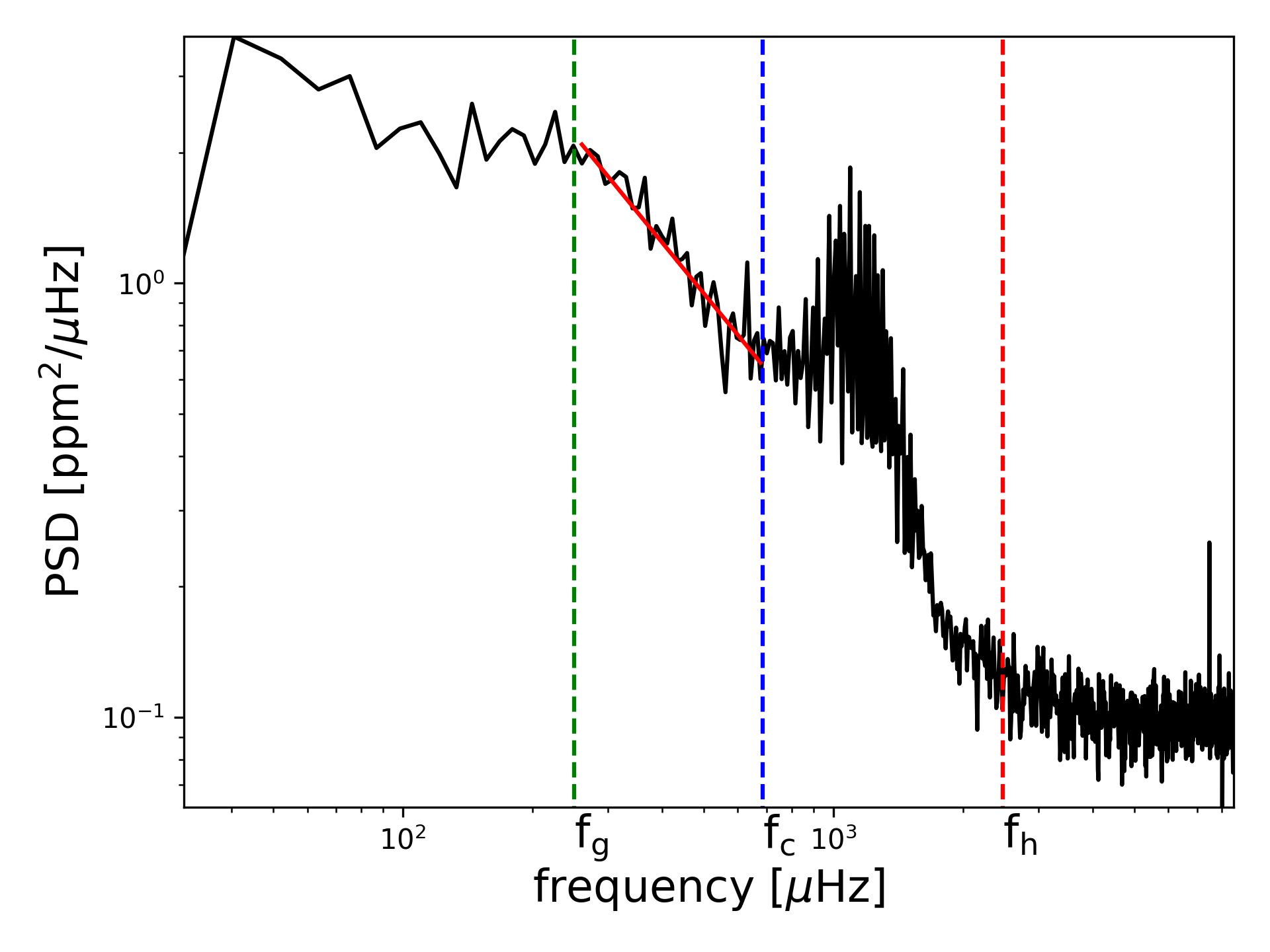}}
    \caption{Example of an averaged periodogram of the Kepler target KIC 7940546 ($M_s = 1.152~M_\odot$, $R_s = 1.807~R_\odot$, $T_{eff} = 6244~K$, 
$m_{v} = 7.397$). The number of available one-day subseries for this target is 
$L=75$. Cut-off frequencies $f_c$ and $f_g$ resulting from the MCMC analysis are indicated by the vertical dashed lines.}
    \label{Fig_Kep1}
\end{figure}

\subsection{Frequency domain analyses}
\label{Sec42}

For each target, we selected the whole number of available one-day subseries ensuring a minimum number of data points per series. 
The required condition for a subseries to be considered is that it must have no more than $10\%$  missing data, with a maximum length for a single gap of $five$ consecutive data points. Using these subseries, we then generated the corresponding averaged periodogram using Eq. \eqref{eq_PL} for each target.  
An example of one of these periodograms for an F6IV star is shown in Fig.~\ref{Fig_Kep1}.

As in Sec.~\ref{Sec2}, we aim to describe these stellar PSDs with power law functions defined as in Eq. \eqref{eq_DSP}. 
For this purpose, the target PSDs have to be split into four regions that are: the high-frequency region (A) delimited by $f_h$, 
the p-mode region (B) delimited by $f_h$ and $f_c$, the flicker region (C) delimited by $f_c$ and $f_g$, and the low-frequency region (D) delimited by $f_g$ 
and the frequency cut-off of the SG filter, $f_\ell$. These regions are labelled in Figs.~\ref{Fig_carac} and \ref{Fig_Kep1}.
To improve the automatic identification of these cut-off frequencies, and because the Kepler observations are noisier than solar VIRGO observations, 
we made use of \textit{a priori} knowledge on the mean oscillation frequency ($\nu_{max}$) to help the localisation of the oscillation p-modes (which are not clearly visible in all Kepler periodograms). Following \citetads{1991ApJ...368..599B}, \citetads{1995A&A...293...87K}, and \citetads{Belkacem_2011}, we estimated $\nu_{max}$ based on the stellar parameters as:
\begin{equation}
\nu_{max} = \nu_{max, \odot} \times \Big(\frac{M_s}{M_\odot}\Big) \times 
\Big(\frac{R_s}{R_\odot}\Big)^{-2} \times 
\Big(\frac{T_{eff}}{T_{eff,\odot}}\Big)^{-0.5},
\label{eq_numax}
\end{equation}
with $\nu_{max, \odot} = 3150 ~\mu$Hz being the solar value.
Moreover, we fixed frequency $f_h$ to the solar VIRGO value (i.e., $4000.15$ $\mu$Hz, see Sec.~\ref{Sec2}) as we found this value to be a good estimate for all considered stars with detectable acoustic modes signatures (mostly G and F stars).
For each star, we then inferred the parameters of Eq. \eqref{eq_DSP} and the cut-off frequencies using MCMC analyses. For each MCMC, the fitted (``jump'') parameters are the two cut-off frequencies $f_c$ and $f_g$, the three associated power indices $\mathbf{\alpha} := [\alpha_{h}, \alpha_{c}, \alpha_{g}]^T$, and constants
$\mathbf{\beta} := [\beta_{h}, \beta_{c}, \beta_{g}]^T$ as defined in Eq. \eqref{eq_DSP}.
The starting values were set to the solar values obtained in Sec.~\ref{Sec2} and no other prior than $\nu_{max}$ was used to avoid influencing the results.

\begin{figure}[t] \centering
    \resizebox{\hsize}{!}{\includegraphics[width=\linewidth]{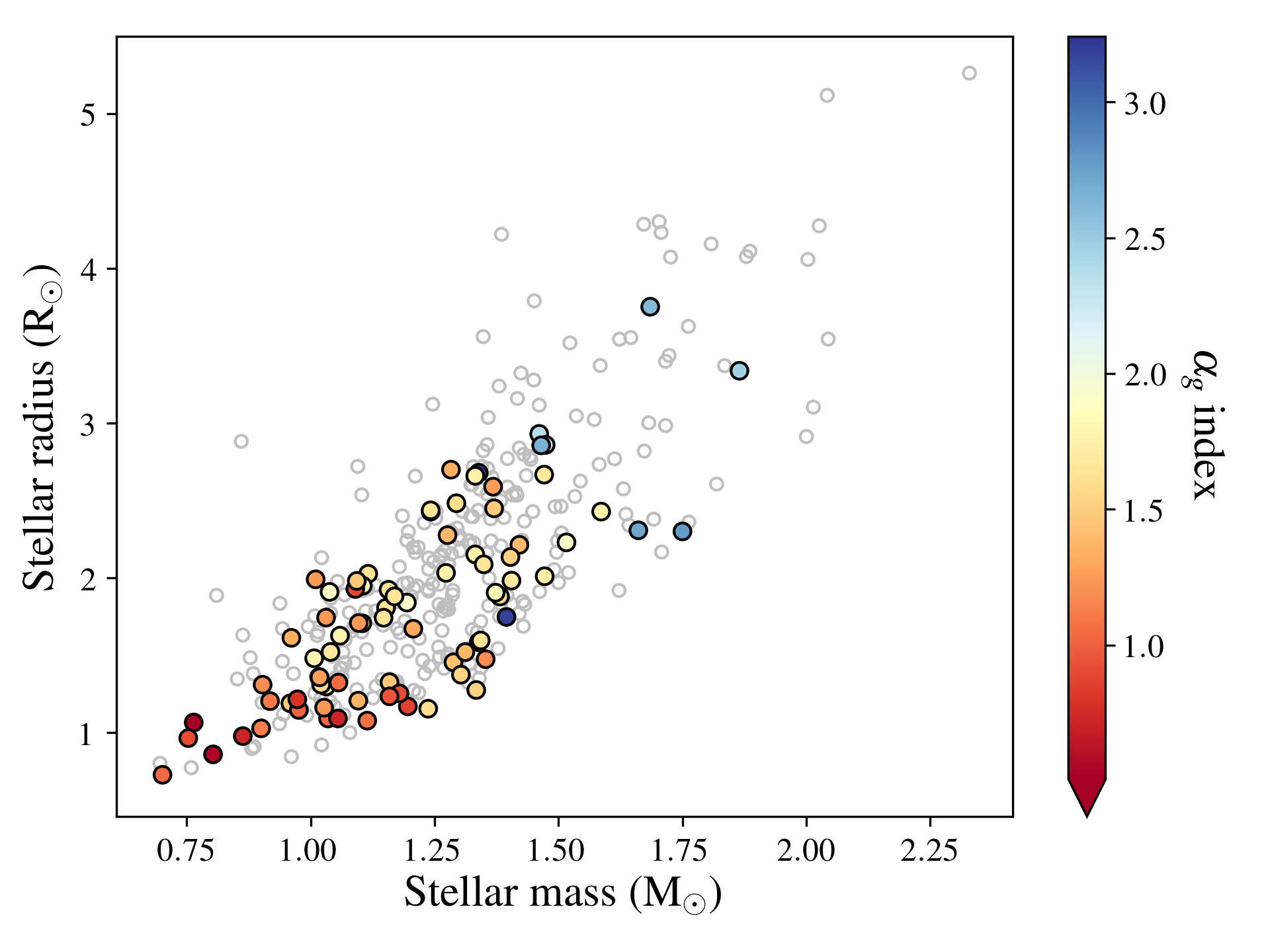}}
    \caption{Selected Kepler targets represented as a function of their mass and 
radius. The color code indicates the inferred index parameters in the frequency 
region of granulation ($\alpha_g$) derived after correction by the HF noise. The best targets with $R^2>0.9$ are shown in color (see Sec.~\ref{Sec43}).}
    \label{Fig_Kep0}
\end{figure}

\begin{figure*}
    \resizebox{\hsize}{!}{\includegraphics{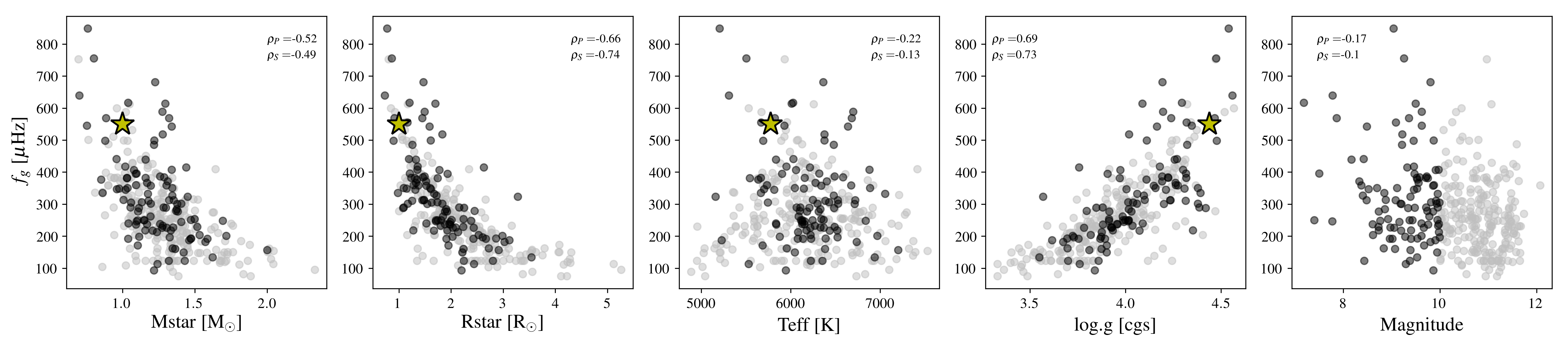}}
    \caption{Estimated values of the cut-off flicker frequency $f_g$ as a function of stellar parameters resulting from the MCMC analyses performed on periodograms of the selected SC Kepler targets. From left to right: stellar mass, radius, effective temperature, surface gravity, and apparent magnitude.
    Black dots represent targets with magnitude $m_v \le 10$ and gray dots targets with magnitude $m_v>10$. The star symbol shows the solar cut-off frequency evaluated from VIRGO data (see Sec.~\ref{Sec_24}). Pearson ($\rho_P$) and Spearman's ($\rho_S$) coefficients, evaluated for stars with $m_v \le 10$, are indicated in each panel. }
    \label{Fig_Kep2}

    \resizebox{\hsize}{!}{\includegraphics{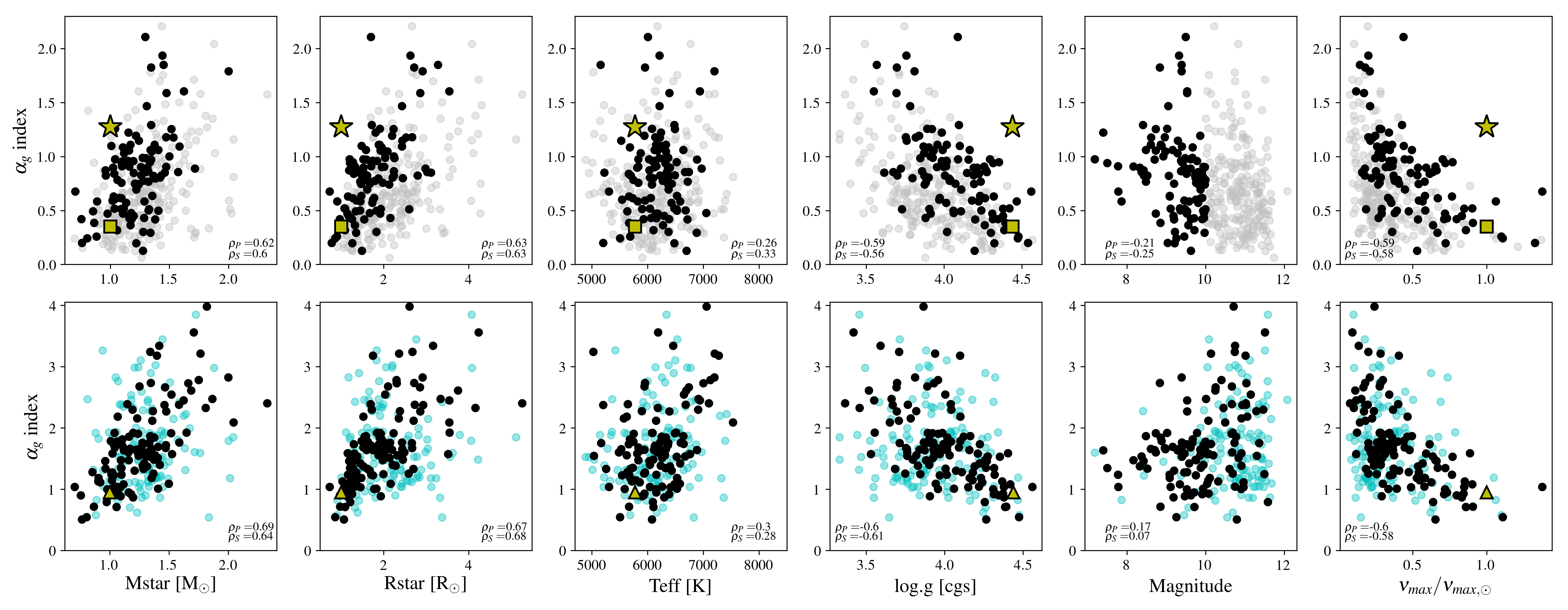}}
    \caption{Estimated flicker index associated to granulation ($\alpha_g$) as a function of stellar parameters. From left to right: stellar mass, radius, effective temperature, surface gravity, apparent magnitude, and normalized $\nu_{max}$ resulting from \eqref{eq_numax}.
    Top: Values obtained from the MCMC analyses. The color code indicates the apparent magnitude of the target: $m_v<10$ (black) and $m_v>10$ (gray). The star 
symbol represents the index derived for the Sun based on VIRGO green channel observations: $\alpha_g=1.26$ with $\sigma_W=5$ ppm.  The square symbol 
represents the index obtained after adding a HF noise level (corresponding to that seen in Kepler observations of the Sun-like star KIC 3427720) to the VIRGO subseries. Bottom: Values obtained after interpolation that corresponds to a HF level 
of $\sigma_W=30$ ppm (see Sec.~\ref{Sec43}). The color code indicates here the 
data with $R^2>0.8$ (i.e., the best fits, black) and $R^2>0.5$ (blue).
   The triangle symbol represents the value of the flicker index derived by 
adding a WGN of $\sigma_W=30$ ppm in solar VIRGO observations, for which we 
find: $\alpha_g=0.9464$. This is the raw level of HF noise we expect for 
Sun-like stars observed with CHEOPS. 
   Pearson ($\rho_P$) and Spearman's ($\rho_S$) coefficients associated with these plots for the best targets ($R^2>0.9$) are indicated on each panel. }
    \label{Fig_Kep3}

    \resizebox{\hsize}{!}{\includegraphics{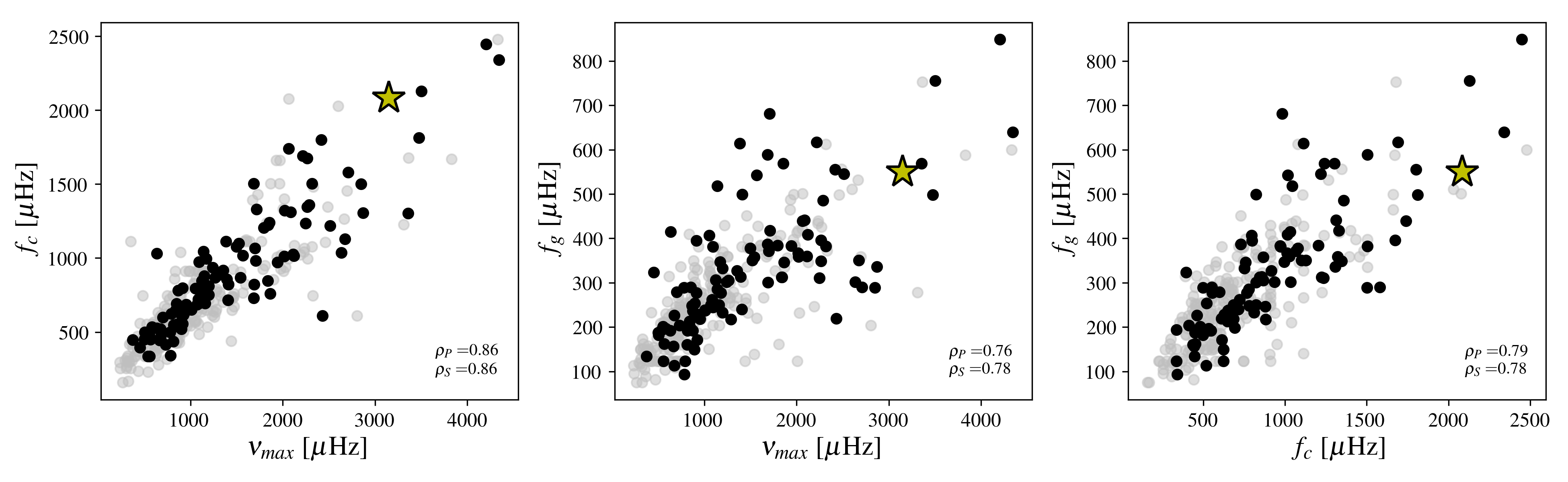}}
    \caption{Correlations between $\nu_{max}$ (determined using Eq. \eqref{eq_numax}), the corner frequency, and the flicker frequency. The corner 
and flicker frequencies have been derived for each Kepler target using the MCMC 
analysis described in Sec.~\ref{Sec42}. The color code indicates the apparent 
magnitude of the target: $m_v<10$ (black) and $m_v>10$ (gray). Solar values 
derived from VIRGO observations are shown by the yellow star in each panel. Pearson ($\rho_P$) and Spearman's ($\rho_S$) coefficients associated with these plots are indicated on each panel.}
    \label{Fig_Kep3b}
\end{figure*}

For approximately two-thirds of the targets in our sample, we found a flicker index $\alpha_g<0.1$ with large uncertainties, which means that the PSD slope in the frequency range of granulation is not clearly detected. 
As granulation signals are undetectable for these targets, we do not consider them in the following analysis. 
Our final set of targets contains $335$ stars, among which we selected the $82$ ``best'' targets, as these are bright ($m_v<10$, see Figs.~\ref{Fig_Kep0}), which makes the measurement of the granulation parameters ($f_c$, $f_g$ and $\alpha_{g}$) more precise. 
The values for $f_g$ and $\alpha_{g}$ resulting from the MCMC analyses of our $335$ targets are shown as a function of the stars' fundamental parameters\footnote{taken in the \textit{Kepler$\_$stellar17.csv.gz} catalog, \url{https://archive.stsci.edu/kepler/catalogs.html}} in Figs.~\ref{Fig_Kep2} and \ref{Fig_Kep3} (top panel).

In both figures, the subsample of best targets is highlighted. 
Moreover, we quantified the strength of any potential correlation between the fitted parameters ($f_g$ and $\alpha_{g}$) and the stellar parameters using the Pearson ($\rho_P$) and Spearman's ($\rho_S$) coefficients (see numerical values in each panel).

As shown in Fig.~\ref{Fig_Kep2}, we observe strong correlations (i.e., $|\rho_P|, |\rho_S|>0.2$) between the flicker frequency $f_g$ and the stellar mass, radius, and surface gravity. These correlations are particularly noteworthy when placing the solar $f_g$ value derived from VIRGO data on these plots 
(shown as star symbols). As expected, the granulation timescales decrease, that is, the flicker frequencies increase, for decreasing stellar mass and radius.
This is in agreement with the work of \citetads{2011ApJ...741..119M}, who extracted the characteristic timescales of granulation based on different Harvey law fits on Kepler red giant stars.
We note that no significant correlation is observed with the stellar magnitude ($|\rho_P|, |\rho_S| \sim 0.1$), which means that the flicker frequency can be derived independently of the HF noise level for all targets for which granulation as a whole is detectable (i.e., $\alpha_{g}>0.1$). 

In Fig.~\ref{Fig_Kep3b}, we also show the characteristic frequencies $f_c$ and $f_g$ as a function of 
the oscillation frequency $\nu_{max}$ that has been derived using Eq. \eqref{eq_numax}. 
We observe clear linear correlations between these three stellar characteristic frequencies. This indicates that the typical granulation timescales, combined with the mean frequency of the acoustic modes, may be able to track stellar characteristics, such as the stellar surface gravity \citepads{Kallingere1500654}.

In Fig.~\ref{Fig_Kep3} (top panel), we display the inferred power index $\alpha_{g}$ resulting from the MCMC analyses as a function of the stellar parameters. 
As for parameter $f_g$, we observe strong correlations between this parameter and stellar mass, radius, surface gravity, and acoustic oscillation frequency $\nu_{max}$ (see last column). However, we also observe a significant correlation with stellar apparent magnitude.
This indicates that the inferred flicker indices are influenced by the high level of HF noise, which biases the correlations seen with the stellar parameters.
This is particularly evident when comparing the inferred $\alpha_g$ with the flicker index extracted from VIRGO solar observations (which have very low HF noise $\sigma_W\sim5$ ppm, shown with stars symbols).
To compare the solar flicker index with Kepler observations, and to coherently interpret Kepler data, we need to find a way to combine flicker indices computed from PSDs with different white noise levels. 
For this purpose, we added different levels of synthetic WGN $\sim {\cal{N}}(0,\sigma^2_{W})$ with variance $\sigma^2_{W}$ to the solar data and computed the flicker indices of these data as done for the Kepler observations.
Figure~\ref{Fig_Kep4} shows the fast decrease of the solar flicker index with increasing HF noise ($\sigma_W$). 
The black dots represent the flicker index measured on the averaged periodogram of several Sun-like stars observed by Kepler as a function of the level of HF noise (see list in Table.~\ref{Table2}). These values illustrate that correlated noise due to granulation becomes more difficult to detect with increasing HF noise. Uncertainties on $\alpha_g$ are directly related to the number of one-day subseries available to compute the averaged periodogram.
When correcting the measured solar flicker index in Fig.~\ref{Fig_Kep3} (star symbols) for the effect of additional WGN corresponding to that present in Kepler observations, we found a solar flicker index in better agreement with the correlations observed between the Kepler stars and the stellar parameters (see square symbols).

However, to derive the correct relation between the flicker index and the stellar parameters, we have to account for different levels of HF noise when comparing the derived flicker indices. This is the objective of the following section.
 
\subsection{Flicker index derived at a constant HF noise level}
\label{Sec43}

\begin{figure}[t] \centering
    \resizebox{\hsize}{!}{\includegraphics{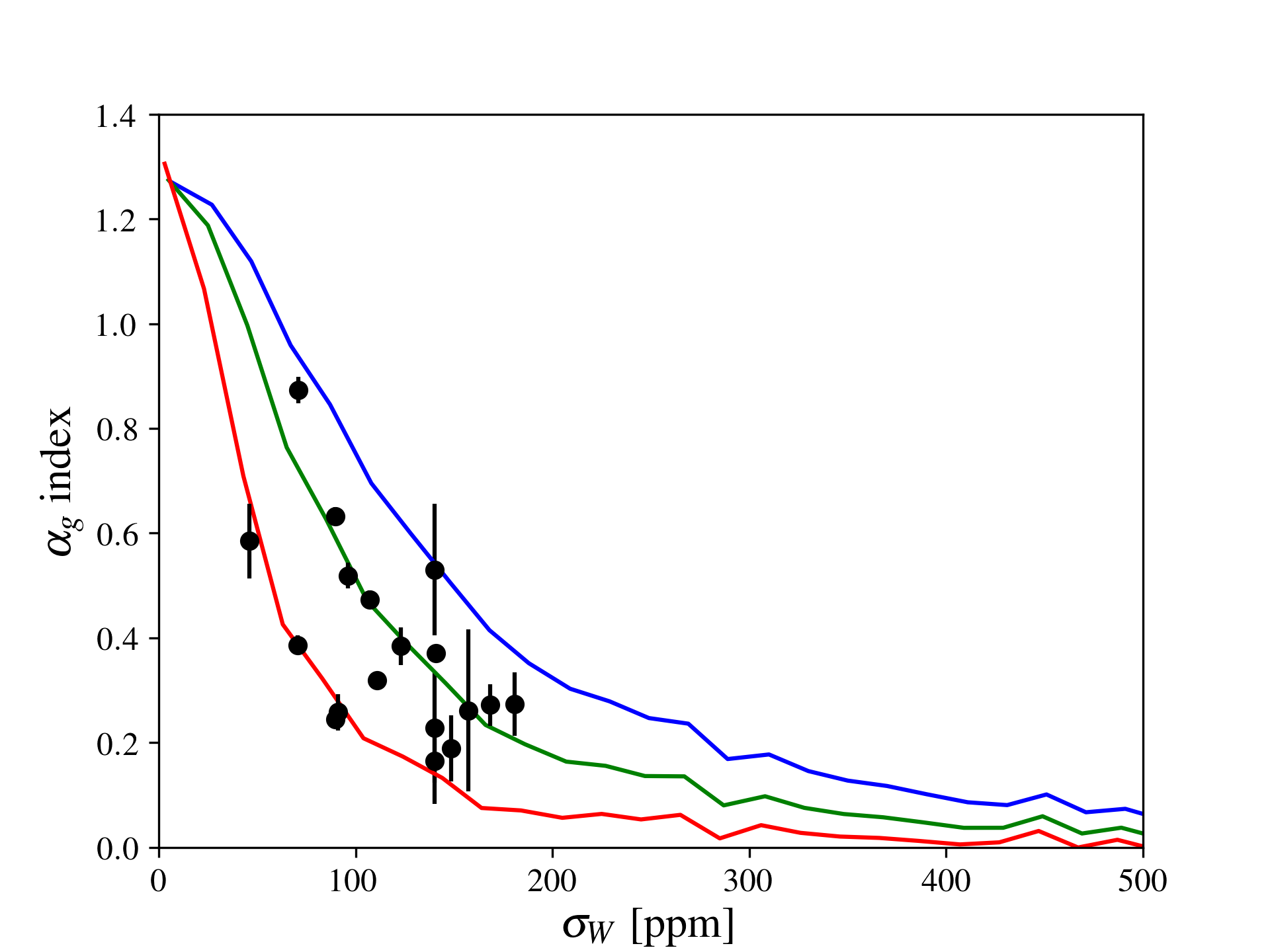}}
    \caption{Estimated values of the flicker index as a function of the HF noise level added in the VIRGO time series (red, blue and green SPM channels). Symbols show the power index measured on the PSD of Kepler Sun-like stars listed in Table.~\ref{Table2}.}
    \label{Fig_Kep4}
\end{figure}

To correct for the influence of the HF noise level on the flicker index and derive unbiased correlations between $\alpha_g$ and the stellar parameters, 
we choose to rely on interpolation techniques. 

\begin{figure}[t] \centering
    \resizebox{\hsize}{!}{\includegraphics{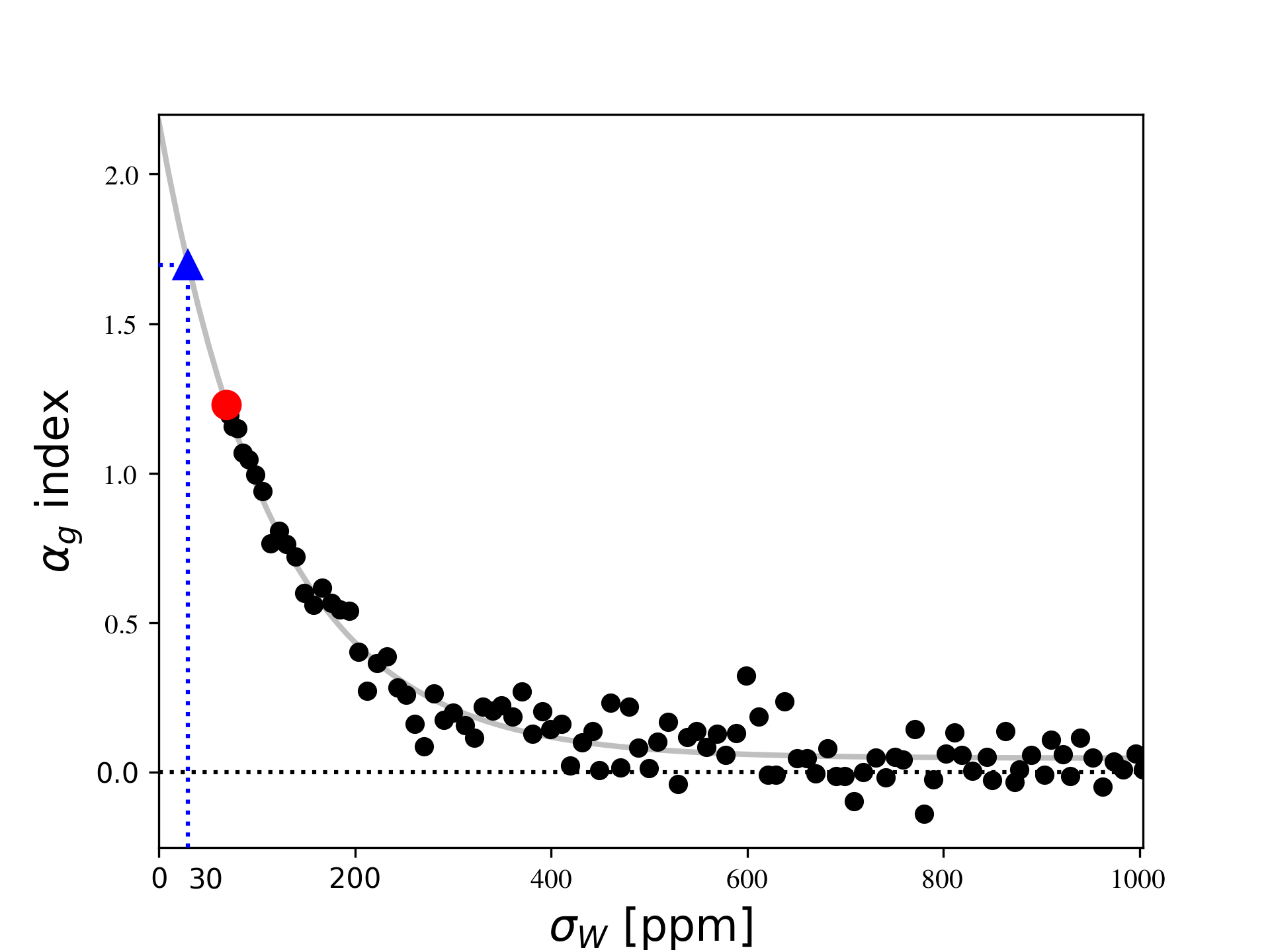}}
    \caption{Flicker index $\alpha_g$ as a function of the level of HF noise (black) for the F-star KIC 7940546. The first value, computed from raw Kepler observations, is indicated by the red dot and corresponds to $\alpha_g=1.23$ and 
$\sigma_W=69$ ppm. The gray line shows the interpolated function (see Eq. \eqref{eq_expo}), for which we obtained a quality factor of $R^2=0.96$. The interpolated index at 
$\sigma_W=30$ ppm is $\alpha_g=1.69$ (blue triangle). }
    \label{Fig_Kep6}
\end{figure}

For each Kepler target, we first empirically measured the decrease of the power index as a 
function of added HF noise  ($\sigma_W$).
We proceed in the same way as for VIRGO solar observations (see Sec.~\ref{Sec42} and Fig.~\ref{Fig_Kep4}). 
For each considered value $\sigma_W$, we added a synthetic WGN $\sim {\cal{N}}(0,\sigma^2_{W})$ to the available one-day subseries, computed the averaged periodogram, and derived the flicker index associated to the frequency region $\nu\in[f_g, f_c]$, with $f_g$ and $f_c$ the flicker cut-off frequencies of the star in question (see Sec.~\ref{Sec42}). 
An example is shown in Fig.~\ref{Fig_Kep6} for a bright  F-star (black dots). For this particular target, we measured an initial HF noise level of $\sigma_W=69$ ppm that corresponds to $\alpha_g=1.23$ (red dot). 
We then used the measured decrease to extrapolate the flicker index towards smaller values, using an exponentially decreasing function of the form:
\begin{equation}
\alpha_g(\sigma_W) = a ~ e^{-b~\sigma_W} + c,
\label{eq_expo}
\end{equation}
with parameters $\{a,b,c\}\in\mathbb{R}$ to be fitted.
We performed a least-square regression of our empirical curve $\alpha_g(\sigma_{W})$ to derive the coefficients $\{a,b,c\}$ (gray line in Fig.~\ref{Fig_Kep6}). 
For this example target, we found a corrected power index of $\alpha_g=2.1$ at the HF noise level of the solar VIRGO observations (i.e., $\sigma_W=5$ ppm). 

We performed similar least-square regressions for all targets in our sample. As expected, the quality of our interpolations depends on the initial level of HF noise that is present within the data: the higher the HF noise, the more inaccurate the interpolated flicker index at low  $\sigma_{W}$. 
To measure the goodness-of-fit, we used the \textit{coefficient of determination} (also known as r-squared coefficient $R^2$), that gives an idea of the distance between the best fit and the observed data points \citepads{doi:10.1002/bimj.19620040313}. 

\begin{figure*}[t]
    \resizebox{\hsize}{!}{\includegraphics{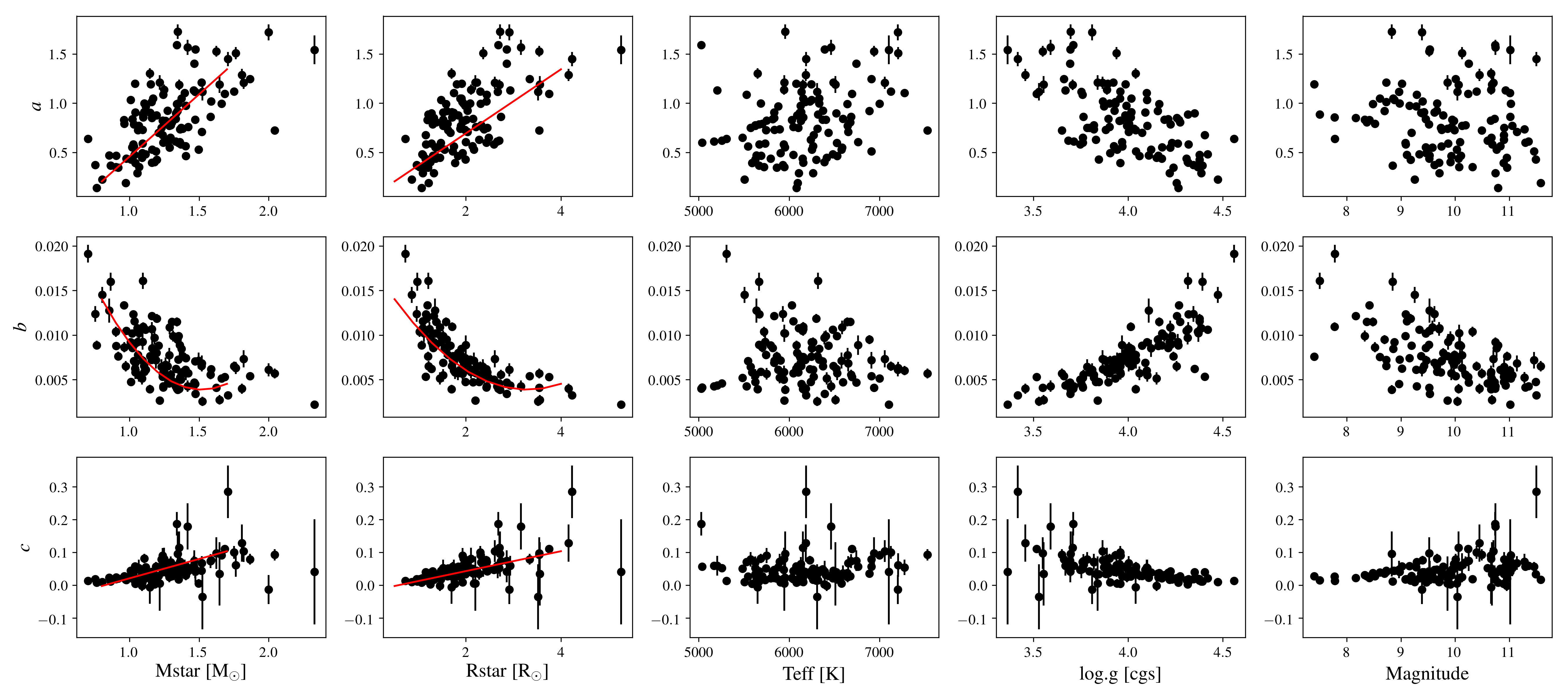}}
    \caption{Coefficients $\{a,b,c\}$ involved in Eq. \eqref{eq_expo} as a function of the Kepler stellar parameters. Red curves represent the linear and quadratic functions described in Eq. \eqref{eq_abcd}.}
    \label{Fig_Kep7}
\end{figure*}

We disregarded all targets with a corrected power index with $R^2<0.5$, leaving us with $245$ Kepler targets with $R^2>0.5$, among which $118$ have $R^2>0.8$. The corresponding parameters $\{a,b,c\}$ associated with model Eq. \eqref{eq_expo} for these remaining targets are shown as a function of the stellar parameters in Fig.~\ref{Fig_Kep7}. We observe strong correlations between these parameters and the stellar parameters. The combination of these correlations with Eq. \eqref{eq_expo} allows us to estimate the flicker index we will observe for a target star observed with a given level of HF noise and make some predictions for future high-precision observations of CHEOPS and PLATO (see Sec.~\ref{Sec5}).

We then chose a reference level of $\sigma_{W}=30$ ppm, because this is close to the HF noise level expected to be reached with CHEOPS for Sun-like stars with a magnitude $m_v<8$ (see following Section~\ref{Sec5}). The corrected indices (interpolated at the level of $\sigma_{W}=30$ ppm) are shown as a function of the stellar parameters in the bottom panel of Fig.~\ref{Fig_Kep3}. Comparing with the raw flicker indices of the top panels, we observe more significant correlations with the stellar parameters, as expected. The Pearson and Spearman's coefficients reveal strong positive correlations with the stellar mass and radius and negative correlation with the surface gravity, in particular when considering only the best targets ($R^2>0.9$, black dots). If we include the whole Kepler sample (i.e., $0.5<R_2<1$, gray dots), the correlations are slightly less pronounced as the flicker indices show a larger dispersion. We expect these relations to become increasingly precise with future high-precision observations of CHEOPS and PLATO.

%%%%%%%%%%%%%%%%%%%%%%%%%%%%%%%%%%%%%%%%
% SEC.V: Prediction for CHEOPS observations
%%%%%%%%%%%%%%%%%%%%%%%%%%%%%%%%%%%%%%%% 
%\newpage
\section{Predictions for CHEOPS and PLATO}
\label{Sec5}

CHEOPS is the first ESA S-class mission. Its objective is to characterize 
transiting extrasolar planets with high-precision photometric observations 
\citepads{2014SPIE.9143E..2JF}. The passband ($\lambda \in [400,1100]$ nm) and 
high cadence ($1$ min) of CHEOPS will be similar to the Kepler SC observations. 
However, CHEOPS will mainly focus on bright stars making this instrument a very 
promising tool to characterize the stellar variability affecting high-precision observations. For example, \citetads{2018A&A...620A.203M} recently analyzed the 
detectability of the oscillation frequency $\nu_{max}$ on main sequence bright stars that will be observed with CHEOPS. These latter authors found that $\nu_{max}$ will be detectable on most main sequence stars, which can help to precisely 
constrain the age, mass, radius, and density of the host stars, also aiding the characterization of the observed transiting planets. 

In this section, we explore to what extent the upcoming missions CHEOPS and PLATO will allow us to characterize stellar granulation through the power index defined in Sec.~\ref{Sec2}. 
We consider flicker noise to be detectable in light curves with inferred power indices of $\alpha_g>0.2$, while HF noise dominates otherwise. 
Our objective is to derive the limiting magnitudes ($m_{v,lim}$) for which our measurements possess the necessary precision to measure at least this limiting flicker index $\alpha_{g,lim}$.

In Sec.~\ref{Sec4}, we showed the dependence of this index on the level of HF noise, which is related to the stellar apparent magnitude of the target star. We also derived the relation between the flicker index and the level of HF noise through Eq. \eqref{eq_expo}, which involves a set of parameters $\{a,b,c\}$. These parameters are correlated with the stellar mass ($M_s$) and radius ($R_s$, see Fig.~\ref{Fig_Kep7}). 

In the following, we consider main sequence G- and F-type stars (as well as slightly evolved F-stars) that are known to host a convective envelop. 
Our set of stellar parameters $\{M_s,R_s\}$ encapsulates: 
\begin{itemize}
\item stars with $0.8 M_\odot \le M_s < 1.4 M_\odot$ and $0.8 R_\odot \le 
R_s \le 2.5 R_\odot$, 
\item stars with $1.4 M_\odot \le M_s \le 1.5 M_\odot$ and $1.7 R_\odot \le 
R_s \le 2.5 R_\odot$,
\item stars with $M_s = 1.6 M_\odot$ that have a shallow convective envelope, though still present, at $R_s=2.1$, $2.2$ and $2.3$ $R_\odot$.
\end{itemize}
We do not include stars of M and K spectral types with $M_s<0.8M_\odot$ because the flicker indices of such stars were not constrained by our sample of Kepler observations (see Sec.\ref{Sec4}). Moreover, according to  stellar models based on the \textit{Code Liégeois d’Evolution Stellaire} (CLES) 
stellar evolution code \citepads{2008Ap&SS.316...83S,2019ApJ...879...94F}, no sufficiently thick convective envelopes are expected for stars with masses above $1.6 M_\odot$.

For each set of parameters $\theta_s := \{M_s,R_s\}$, we  first predicted the values of parameters $\{a,b,c\}$ as defined in Eq. \eqref{eq_expo} using 
linear and quadratic functions of the stellar mass and radius. We found the following relations (see red lines in see Fig.~\ref{Fig_Kep7}): 
\begin{equation}
\left\{
\begin{aligned}
    a(\theta_s) &= 0.631 M_s + 0.162 R_s -0.379,\\ 
    b(\theta_s) &= 0.010  M_s^2 - 0.025  M_s + 0.00062 R_s^2 - 0.0057 R_s  +  0.030,\\ 
    c(\theta_s) &= 0.002 M_s + 0.029 R_s -0.019.\\
\end{aligned}
\right.
\label{eq_abcd}
\end{equation}
We then derived the HF noise level ($\sigma_{W,up}$) corresponding to the limiting flicker index $\alpha_{g,lim}$ using Eq. \eqref{eq_expo} with parameters $\{a,b,c\}$ given by  Eq. \eqref{eq_abcd}. 
For each target in our defined $\{M_s,R_s\}$ grid, we came up with the highest HF noise level, $\sigma_{W,up}$, that is acceptable to observe a flicker index $\alpha_g\ge\alpha_{g,lim}$. 
We then derived the limiting magnitudes $m_{v,lim}$ corresponding to $\sigma_{W,up}$.  
 To do so, we computed the expected precision for CHEOPS under the assumption that the targets will be observed during times when stray light from the Earth is no more than $0.62\, \mathrm{phot\,.s^{-1}\,.pix^{-1}}$, a medium value found from simulation, accounting for all other noise sources as done in the mission's Exposure Time Calculator\footnote{available from \url{https://www.cosmos.esa.int/web/cheops-guest-observers-programme/ao-1}}. We used a $5920$ K black-body SED (similar to that of a G0 star, typical for our sample), and a time window of one hour.  

However, it is important to note that the number of one-day subseries ($L$)  also plays a role as the variance of the averaged periodogram at a given frequency $\nu$ decreases with $L$ (and so do the uncertainties on $\alpha_g$). 
 
\begin{figure*}[t] \centering
    \resizebox{\hsize}{!}{\includegraphics{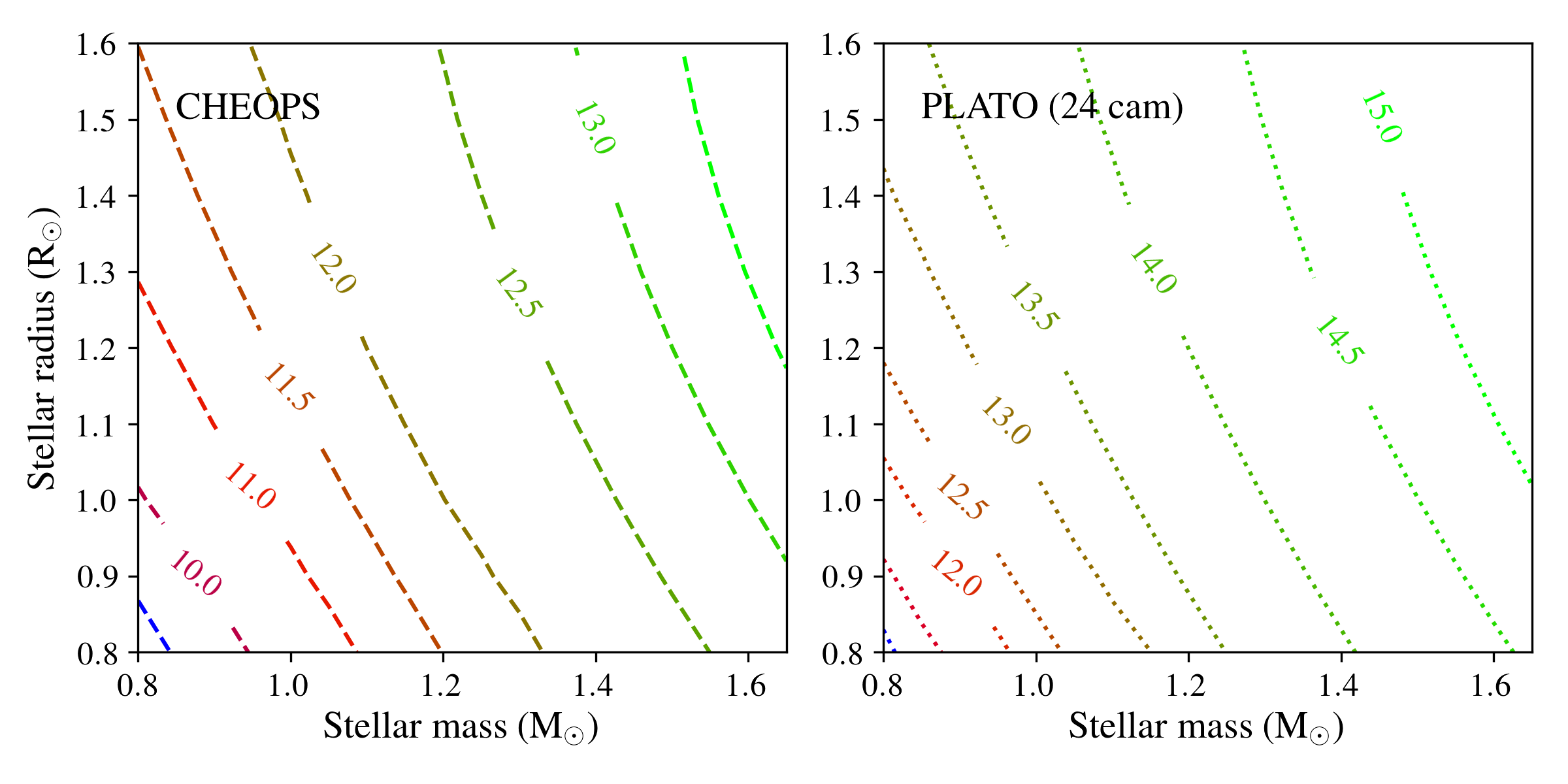}}
    \caption{Illustration of the limiting stellar apparent magnitude, depending on the stellar parameters, (radius and mass) that is needed to measure a flicker index with $\alpha_g>0.2$ with the future CHEOPS (left) and PLATO (right) high-precision observations. }
    \label{Fig_CHEOPS1}
\end{figure*}

The limiting magnitudes (i.e., those corresponding to $\sigma_{W,up}$) derived for the grid of $\{M_s,R_s\}$ are shown in the left panel of Fig.~\ref{Fig_CHEOPS1}. For solar-like stars, flicker noise will become relevant for magnitudes brighter than $m_v\le 10$. 
This encapsulates a large fraction of the expected targets observed with CHEOPS. However, for slightly evolved F-stars, the limiting magnitude will be 
around $13$ and therefore flicker noise is expected to be relevant in most if not all CHEOPS light curves of bright F-type stars. For these stars, precise flicker index values should be measurable.

To make predictions for the PLATO mission 
\citepads{2014ExA....38..249R}, we use the precision estimate by \citetads{2014A&A...566A..92M}, who (when using $24$ cameras) quote an expected noise level of $27$, $34,$ and $80$ ppm per hour for stars with $m_v=10.8, 11.3$ and $13$, respectively. 
We show the limiting stellar magnitudes expected from future PLATO observations in the right panel of Fig.~\ref{Fig_CHEOPS1}. We predict a higher impact of flicker noise in PLATO light curves, and expect that the characterization of the flicker properties (amplitudes and timescales) should be well feasible for most F and G stars.

We see through the analysis of solar observations (see Sec.~\ref{Sec3}) that flicker variability can lead to significant errors on the inferred transit 
parameters of the smallest planets in the case of a single (or a small number of) observed transit(s). 
The development of accurate noise modeling procedures will not only allow us to decrease the errors on the transit parameters, but also to strengthen our understanding of the underlying link between the noise properties and stellar physics (see Sec.~\ref{Sec4}). 

%\newpage

%%%%%%%%%%%%%%%%%%%%%%%%%%%%%%%%%%%%%%%%
% CONCLUSIONS
%%%%%%%%%%%%%%%%%%%%%%%%%%%%%%%%%%%%%%%%
\section{Conclusions}
\label{ccl}

We present a statistical characterization of the short-timescale stellar variability associated (mainly) with granulation noise. Based on solar observations, we find this noise source to be: (i) stochastic, (ii) colored, (iii) stationary with respect to the solar cycle, and (iv) wavelength dependent. 
It can generate variability of several hundred parts per million in amplitude.
In the relevant frequency region of the PSD, the introduced correlations can be modeled by a simple power law.
We chose to use the power index resulting from fits to the PSD as an indicator of the noise correlations. We used HMI images of the Sun to create artificial transit light curves of hypothetical planets transiting the Sun, and analyzed the impact of this flicker noise on the inferred transit parameters. We showed that flicker noise is critical for the smallest planets, for which we find that the inferred parameters can be substantially offset from their true values. This however is likely due to inaccurate limb-darkening parameters which have been shown to introduce biases of the same magnitude by \citeads{2016MNRAS.457.3573E}.

We then turned to Kepler short-cadence observations to extract the dependence of the 
flicker power index on the stellar parameters. We found the inferred power index 
values to be heavily affected by the level of high-frequency noise (which is 
related to the stars' apparent magnitude).
Correcting for this influence, we observe a strong correlation between the corrected indices and the stellar radius, mass, and surface gravity. No clear correlation is observed with stellar effective temperature. These correlations confirm the already known relation between this stellar variability and the stellar properties (e.g., see \citeads{2016ApJ...818...43B}).

Using this interpolated power index and the observed dependence with the stellar parameters, 
we predicted the limiting stellar apparent magnitude for which flicker noise will 
be characterizable with future high-precision observations of CHEOPS and PLATO. 
We find that the signature of this noise will be observable for most of the CHEOPS and PLATO light curves of Solar-like targets. 
This study highlights the need to design robust signal processing routines adapted to the characteristics of flicker noise in order to reduce errors on the inferred parameters of small exoplanets, which will be the objective of forthcoming studies.

%%%%%%%%%%%%%%%%%%%%%%%%%%%%%%%%%%%%%%%%
% ACKNOWLEDGEMENTS
%%%%%%%%%%%%%%%%%%%%%%%%%%%%%%%%%%%%%%%%

\begin{acknowledgements}
The authors would like to thank René Heller for his peer review containing very useful suggestions, as well as D. Mary for his constructive comments.
S. Sulis, M. Lendl, L. Fossati and P. Cubillos acknowledge support from the Austrian Research Promotion Agency (FFG) under project 859724 ``GRAPPA''.
V.Van Grootel is a F.R.S.-FNRS Research Associate.
The VIRGO instrument onboard SoHO is a cooperative effort of scientists, engineers, and technicians, to whom we are indebted. SoHO is a project of international collaboration between ESA and NASA.
This paper includes data collected by the Kepler mission. Funding for the Kepler 
mission is provided by the NASA Science Mission directorate.
\end{acknowledgements}

%%%%%%%%%%%%%%%%%%%%%%%%%%%%%%%%%%%%%%%%
% BIBLIOGRAPHY
%%%%%%%%%%%%%%%%%%%%%%%%%%%%%%%%%%%%%%%%
\bibliographystyle{aa} 
\bibliography{bibfile} 

%%%%%%%%%%%%%%%%%%%%%%%%%%%%%%%%%%%%%%%%
% APPENDIX
%%%%%%%%%%%%%%%%%%%%%%%%%%%%%%%%%%%%%%%%

\begin{appendix}

% ---------------------------
\section{Validity of the artificial transit light-curve modeling}
\label{AppA}

\begin{figure}[h!] \centering
    \resizebox{\hsize}{!}{\includegraphics{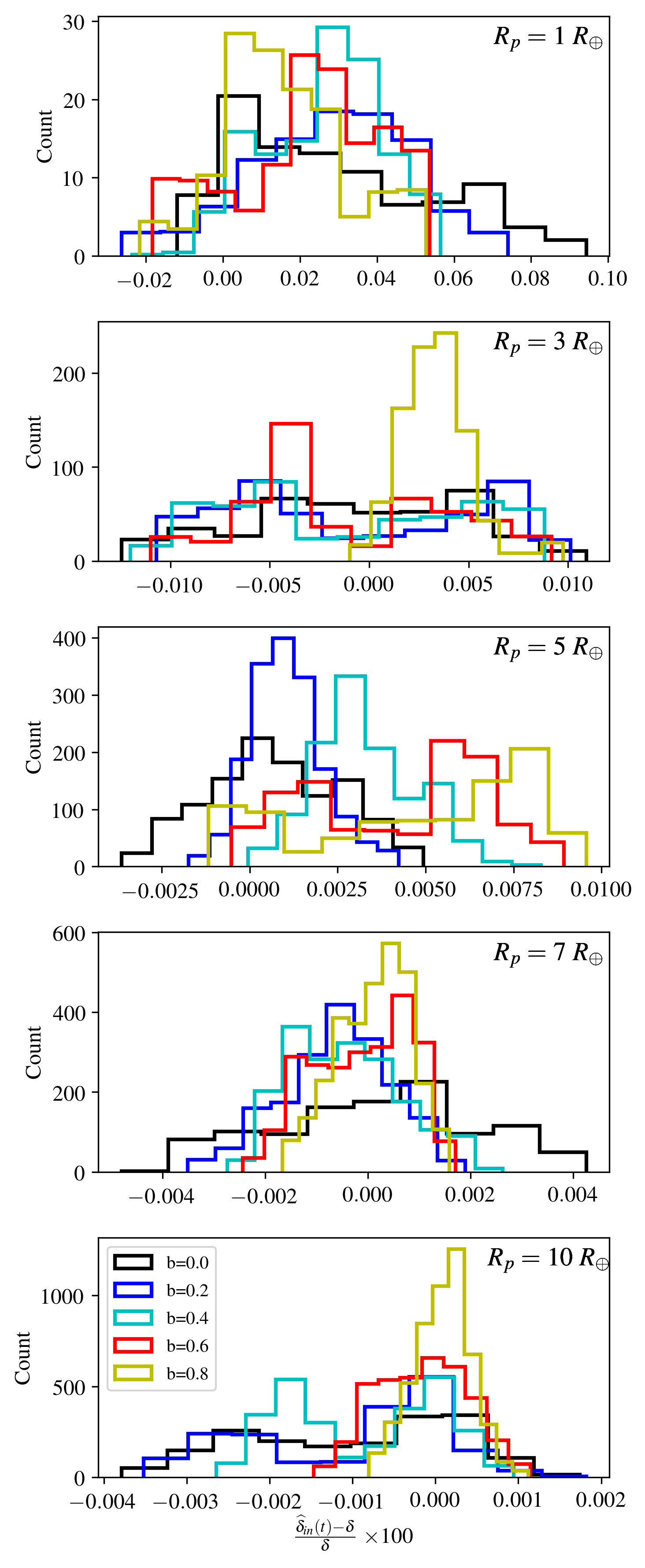}}
    \caption{Distributions of the percentage error on the transit depth ($\delta$) shown for the  artificial transits generated using one solar time series (2018-12-10). Each panel represents the errors for a different planet size ($R_p=1,3,5,7,10$ $R_\oplus$ from top to bottom, resp.) and impact parameter (see legend). The temporal evolution of the ratio of covered to uncovered numbers of pixels has been measured in transit ($\delta_{in}$).}
    \label{Fig_AppA}
\end{figure}

The artificial transit light curve experiment is based on solar observations and 
is subject to unresolved phenomena compared to true extrasolar planet transit 
events: the change of the solar radius over time and the variability of the 
number of the covered pixels by the black body sphere mimicking the planetary 
transit. This leads to temporal variations of the amount of masked solar surface 
that can introduce errors when applying traditional light curve models,
such as those by \citet{2002ApJ...580L.171M}. \\ 
To quantify the error on the transit depth, we computed 
 the ratio of the  sum of the number of pixels covered by the 
exoplanet to the  sum of the number of pixels of the solar disk. Both 
sums evolve as a function of time. We then measured this ratio in-transit 
($\delta_{in}$) and compared it to the true transit depth ($\delta$), taken as 
the square of the planet radius over the solar radius. The distribution of the 
percentage error for each set ($R_p$, $b$) is shown in Fig.~\ref{Fig_AppA}. \\
We observe a percentage error that is $<0.01\%$ of $\delta$ for each set of parameters.
This value is far  smaller than the global error 
found for the inferred transit parameters due to the flicker noise (see Sec.~\ref{Sec32}).
It leads to a difference between the transit model and the artificial transit 
without noise that is $<1$ ppm in-transit and $<2$ ppm in the ingress and egress 
regions of the transits. We note slightly larger differences in the ingress and 
egress regions due to variation of the exact number of pixels covered by the 
planet as a function of time.\\
We conclude that our experiment generating artificial transit light curves in 
solar observations is reliable as the temporal variability of the size of the 
surface area covered by the planet is not significant. This approximation of a 
constant planet-to-sun radius ratio is valid when oversampling the 
raw HMI observations by a factor of two (not shown) or more. In our experiment, we 
oversampled each pixel of the HMI images by a factor of five as we found this 
value to be a good compromise between computational cost and constant number of  
covered pixels over time.

% ---------------------------

\section{Synthetic error bars added to solar HMI observation}
\label{AppB}

Errors are not given for HMI observations but are  necessary information 
to run the MCMC analyses and derive the uncertainties on the inferred transit 
parameters. To add synthetic error bars on our artificial light curve dataset we 
turned to  GP (\cite{Rasmussen:2005:GPM:1162254}).
The GP modeling aims to roughly correct for the 
correlated components of the flicker noise to extract the remaining whitened 
scatter noise. This is a flexible noise modeling commonly used in the exoplanet 
community to take into account the correlated stochastic noise within the 
observations. For that purpose, we use the \textit{George} package developed by 
\citetads{2015ITPAM..38..252A}.

We chose to parametrize the noise covariance matrix in the solar 
observations (without artificial transit) with a product of two kernels that 
\textit{roughly} describe the noise correlation: a constant ($\gamma$) and the 
\textit{Mat\`{e}rn} $3/2$ kernel, the latter being known to be flexible 
regarding unexpected local behavior of the observations and have already been 
used for modeling the short-timescale granulation noise 
\citepads{2018A&A...615L..13G}.
 Explicitly, this kernel writes:
$$
k(\mathbf{x}) = \gamma ~ \Bigg(1+\frac{\sqrt{3}\mathbf{x}}{\ell}\Bigg) 
\exp{\Bigg(-\frac{\sqrt{3}\mathbf{x}}{\ell}\Bigg)},
$$
with $\ell$ being the kernel's metric and $x=|t_i-t_j|$ the data inputs associated to 
the \textit{i}-th and \textit{j}-th data points, respectively.
The covariance matrix, $\mathbf{K}$, is then:
$$
\mathbf{K_{ij}} = \sigma_i^2~\delta_{ij}+k(\mathbf{x}),
$$
with $\sigma_i$ being the uncertainties of the observations at time $i$ and 
$\delta_{ij}$ being the Kronecker delta.

We performed a GP regression following the method described in  
\citet{Gibson12} with the fit quality determined by minimizing the negative 
log-likelihood function corresponding to our GP model:
$$
{\rm log}~{\cal L}(\mathbf{r}) = 
-\frac{1}{2}\mathbf{r}^{\top}~\mathbf{K}^{-1}-\frac{1}{2}{\rm 
log}~|\mathbf{K}|-\frac{N}{2} {\rm log}~(2\pi),
$$
with $\mathbf{r}$ being the data residuals and $N$ the number of data points.

An example of a GP fit is shown in Fig.~\ref{AppB}. 
For each dataset, we measured the standard deviation ($\sigma$) of the data residuals. We obtained $\sigma=20-30$ ppm depending on the considered solar time 
series. We used these error bars in the MCMC simulations described in Sec.~\ref{Sec32}.

\begin{figure}[!h] \centering
    \resizebox{\hsize}{!}{\includegraphics{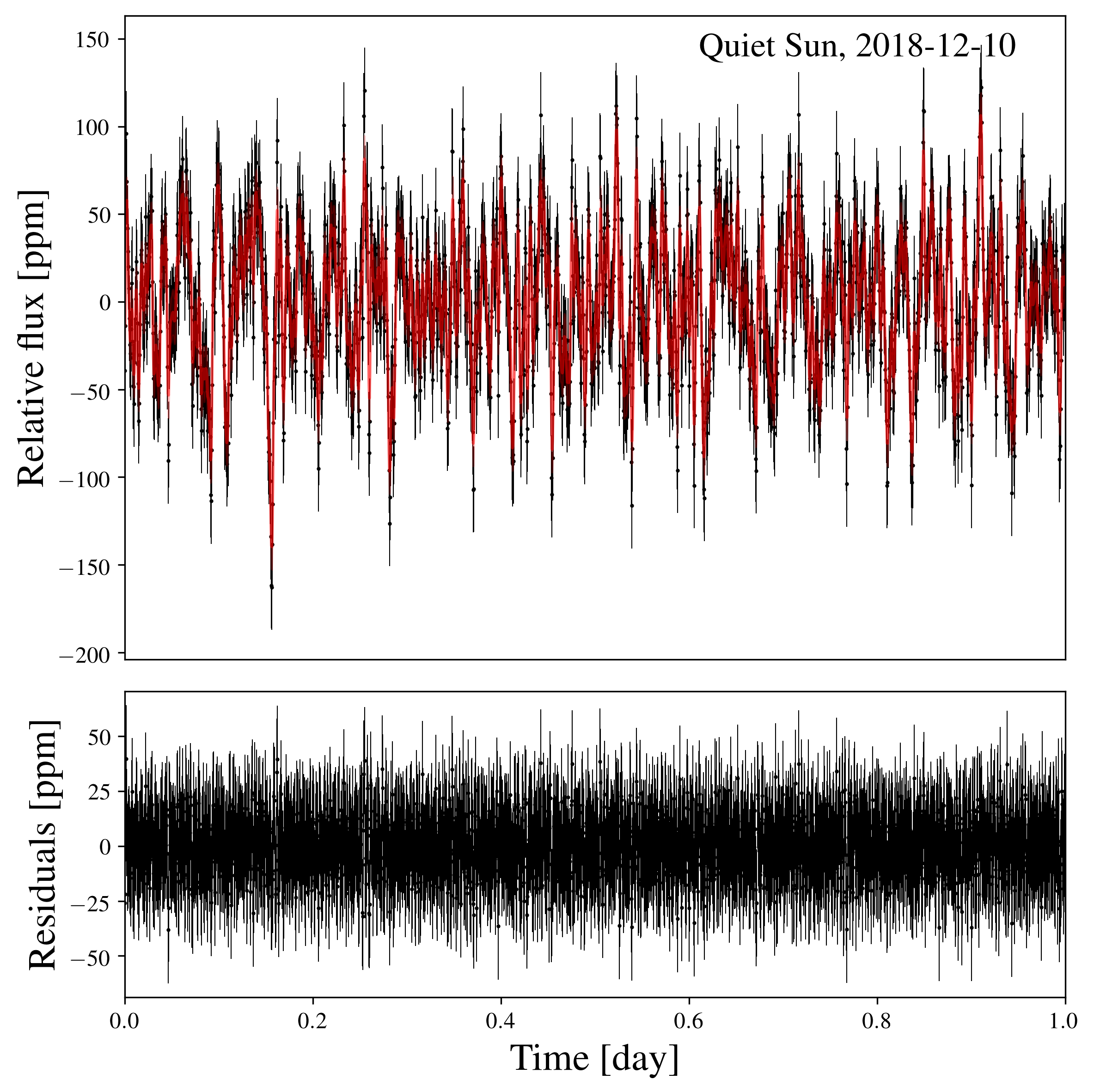}}
    \caption{Top: Example of a raw solar dataset (black) and mean of the 
predictive distribution of the GP model (red). Bottom: Residuals of the solar 
dataset corrected by the GP model. The standard deviation of these residuals is 
used as input synthetic errorbars in our artificial transit light curves. }
    \label{Fig_AppB}
\end{figure}
% ---------------------------

\section{List of Kepler Sun-like stars}
\label{AppC}

\begin{table*}[t]
\centering
\caption{From left to right: KepID, stellar mass, radius, effective temperature, surface gravity, apparent magnitude, number of one-day subseries, level of HF noise, flicker index and associated uncertainties.}
\begin{tabular}{|c|c|c|c|c|c|c|c|c|c|}
\hline
 KepID  &        $M_s$  [$M_\odot$]&    $R_s$  [$R_\odot$]&    Teff  [K]&    logg  [cgs] &    $m_v$  &    $L$   &  $\sigma_W$ [ppm] &  $\alpha_g$ &   $\Delta \alpha_g$ \\
\hline
3427720  &    1.03  &    1.09   &     6045  &     4.37  &     9.10   &    391  &  95.98 &  0.52  &     0.0248 \\
3735871  &    1.05  &    1.09   &     6108   &    4.38   &    9.70  &     493   &  110.95 & 0.32  &     0.0102 \\
5698005  &    0.96  &    0.84   &    5530   &    4.56   &    10.44  &     4    &  140.02 &  0.23  &     0.1049 \\
5724853  &    0.99  &    1.11   &    5959  &     4.34   &    10.25   &    20    & 140.04 &  0.53  &     0.1260 \\
5962180  &    0.88  &    0.90   &    5691   &    4.47   &    9.45   &     54   & 91.17 &   0.26  &     0.0350 \\
6034108  &    0.94  &    1.06   &    5931   &    4.36  &     10.46   &    14   &  148.64 &  0.19  &     0.0630 \\
6116048  &    0.96  &    1.19   &    6031   &    4.26  &     8.41   &     205  & 70.97 &   0.87  &     0.0254 \\
6603624  &    1.03  &    1.16   &    5671   &    4.31  &     9.08   &     332  & 89.60 &   0.63  &     0.0175 \\
7871531  &    0.80  &    0.86   &     5505   &    4.47  &     9.25   &    447  & 89.63 &   0.24  &     0.0150 \\
8394589  &    0.97  &    1.15   &     6147   &    4.30  &     9.52   &    492  &107.06 &    0.47  &     0.0176 \\
8424992  &    0.90  &    1.03   &     5721   &    4.36 &     10.31   &    186  &140.82 &    0.37  &     0.0164 \\
9025370  &    0.86  &    0.98   &     5667   &    4.39   &    8.84   &    182  &70.64 &    0.38  &     0.0187 \\
9410862  &    0.94  &    1.12   &    6046   &    4.316  &    10.71   &    516  &168.32 &    0.27  &     0.0400 \\
10079226 &    1.07  &    1.11   &    5945   &    4.37  &     10.07  &     183  &123.07 &    0.38  &     0.0357 \\
10124866 &    0.88  &    0.91   &    5823   &    4.46   &    7.86   &     51   &46.15 &    0.58  &     0.0713 \\
10215584 &    1.03  &    1.14   &     5893   &    4.33   &    10.60  &    15   &157.37 &    0.26  &     0.1545 \\
10482041 &    1.02  &    0.92   &    5674   &    4.51   &    10.32  &     15   &140.01 &    0.16  &     0.0824 \\
10482869 &    1.08  &    1.00   &    6063   &    4.47   &    10.90  &     19    &180.82 &   0.27  &     0.0608 \\
\hline
\end{tabular}
\label{Table2}
\end{table*}

\end{appendix}

\end{document}